\documentclass[aps,onecolumn,floatfix,superscriptaddress,showkeys,nofootinbib,notitlepage]{revtex4-1}

\usepackage{epsf}
\usepackage{epsfig}
\usepackage{graphicx}
\usepackage{dcolumn}
\usepackage{braket}
\usepackage{bm}
\usepackage{amsfonts}
\usepackage{amsmath}
\usepackage{amssymb}
\usepackage{color,soul}
\usepackage{natbib}
\usepackage[normalem]{ulem}

\newcommand{\tf}[1]{\mathrm{#1}}
\newcommand{\dd}[2]{\frac{\partial {#1}}{\partial {#2}}}
\newcommand{\dt}[2]{\frac{\tf{d} {#1}}{\tf{d} {#2}}}

\newcommand{\tc}[1]{\mathcal{#1}}

\newcommand{\new}[1]{\textcolor{blue}{#1}}
\renewcommand{\new}[1]{{#1}}

\begin{document}

\title{Thermodynamics and steady state of quantum motors and pumps far from equilibrium}

\author{Ra\'ul A. Bustos-Mar\'un}
\email[Electronic  address: ]{rbustos@famaf.unc.edu.ar}
\affiliation{Instituto de F\'isica Enrique Gaviola (CONICET) and FaMAF, Universidad Nacional de C\'ordoba, Argentina}
\affiliation{Facultad de Ciencias Qu\'imicas, Universidad Nacional de C\'ordoba, Argentina}

\author{Hern\'an L. Calvo}
\affiliation{Instituto de F\'isica Enrique Gaviola (CONICET) and FaMAF, Universidad Nacional de C\'ordoba, Argentina}
\affiliation{Departamento de F\'isica, Universidad Nacional de R\'io Cuarto, Ruta 36, Km 601, 5800 R\'io Cuarto, Argentina}

\begin{abstract}
\noindent In this article, we briefly review dynamical and thermodynamical aspects of different forms of quantum motors and quantum pumps. We then extend previous results to provide new theoretical tools for a systematic study of those phenomena at far-from-equilibrium conditions.
We mainly focus on two key topics: (1) The steady-state regime of quantum motors and pumps, paying particular attention to the role of higher-order terms in the nonadiabatic expansion of the current-induced forces. (2) The thermodynamical properties of such systems, emphasizing systematic ways of studying the relationship between different energy fluxes (charge and heat currents, and mechanical power) passing through the system when beyond-first-order expansions are required. We derive a general order-by-order scheme based on energy conservation to rationalize how every order of the expansion of one form of energy flux is connected with the others. We use this approach to give a physical interpretation of the leading terms of the expansion. Finally, we illustrate the above-discussed topics in a double quantum dot within the Coulomb-blockade regime and capacitively coupled to a mechanical rotor. We find many exciting features of this system for arbitrary nonequilibrium conditions: A definite parity of the expansion coefficients with respect to the voltage or temperature biases; negative friction coefficients; and the fact that, under fixed parameters, the device can exhibit multiple steady states where it may operate as a quantum motor or as a quantum pump depending on the initial conditions.
\end{abstract}


\maketitle

\section{Introduction}
\begin{figure}
\begin{center}
	\includegraphics[width=0.8\textwidth]{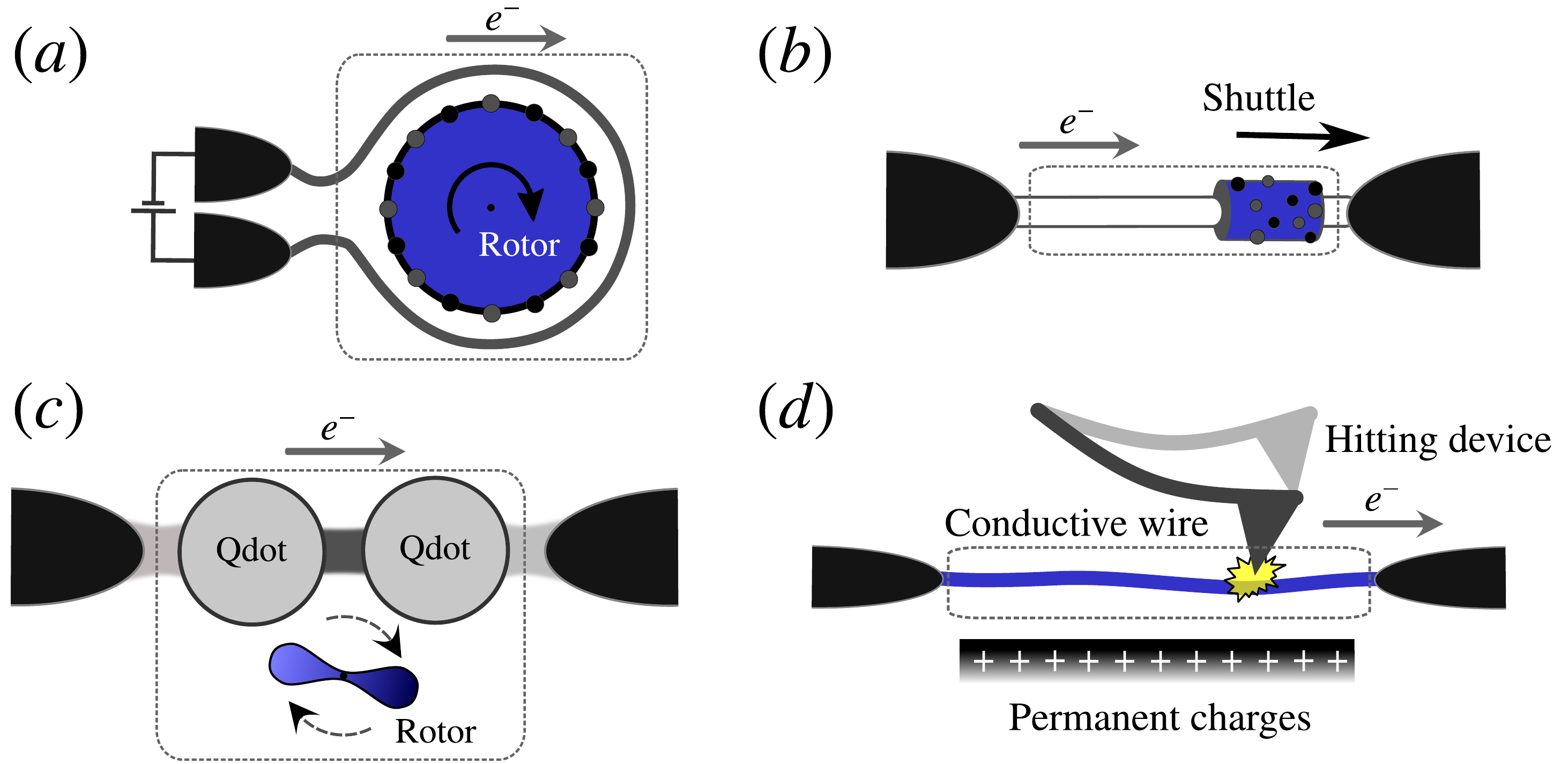}        
\end{center}
\caption{Examples of \textit{local systems} (enclosed by dashed lines) where the movement of a mechanical piece (in blue) is coupled to the flux of quantum particles travelling from/to infinite reservoirs (black
hemiellipses).
(\textbf{a}) A Thouless' adiabatic quantum motor made of charges periodically arranged on the surface of a rotational piece and interacting with a wire coiled around it~\cite{fernandez2017}.
(\textbf{b}) An Anderson's adiabatic quantum motor made of a multiwall nanotube where the outer one, with random impurities, is shorter than the inner one. Another example of it can be made with a rotating piece as in (\textbf{a}), but with charges randomly distributed.
(\textbf{c}) A double quantum dot capacitively coupled to a rotor with positive and negative permanent charges.
The dots are assumed to be weakly coupled to the electron reservoirs~\cite{calvo2017}.
(\textbf{d}) As a result of an external agent, a tip hits a conductive wire capacitively coupled to permanent charges underneath. This starts the oscillation of the wire, which in turn pumps electrons between the reservoirs~\cite{bustos2018}.}                                                                                                                                                                                                                                                                                                                                                                                                                                                                                                                                                                                                                                                                                                                                                                                                                                                                                                                                                                                                                                                                                                                         
\label{fig:1}
\end{figure}

In recent years, there has been a sustained growth in the interest in different forms of nanomachines. This was boosted by seminal experiments~\cite{martinez2016,maillet2019,josefsson2018,kim2014,michl2009,kudernac2011,tierney2011,chiaravalloti2007},
the blooming of new theoretical proposals~\cite{cohen2005,mozyrsky2006,pistolesi2008,cavaliere2009,bustos2013,mazza2014,rossnagel2014,perroni2014,bohrbrask2015,campisi2015,celestino2016,silvestrov2016,romero2017,fernandez2017,bustos2018,bruch2018,rourabas2018,ludovico2018,terrenalonso2019,lin2019,manzano2019,sanchez2019},
 and the latest developments towards the understanding of the fundamental physics underlying such systems~\cite{cohen2003,avron2004,horsfield2004,bennett2010,bode2011,esposito2011,thomas2012,uzdin2015,fernandez2015,uzdin2016,pluecker2017,calvo2017,ludovico2018B,hopjan2018,chen2019}.
Quantum mechanics has proven to be crucial in the description of a broad family of nanomachines which can be put together under the generic name of ``quantum motors'' and ``quantum pumps''~\cite{buttiker1994,brouwer1998,bustos2013,sanchez2016,ludovico2016,ludovico2016B,ludovico2016entropy,benenti2017,whitney2018}. They typically consist in an electromechanical device connected to electronic reservoirs and controlled by nonequilibrium sources, see Figure~\ref{fig:1}. These nonequilibrium sources may include temperature gradients and bias voltages among the reservoirs, or even an external driving of the internal parameters of the system. The dimensions of the electronic component of these devices are normally within the characteristic coherence length of the electrons flowing through them, hence the essential role of quantum mechanics in their description.

Various aspects of quantum motors and pumps have been extensively studied in the literature. For example, it has been shown that quantum interferences can be exploited to boost the performance of these devices. Remarkably, some systems operate solely due to quantum interference, e.g., quantum pumps and motors based on chaotic quantum dots~\cite{cremers2002,bustos2013,fernandez2015}, Thouless quantum pumps and motors~\cite{thouless1983,bustos2013,fernandez2017}, or Anderson quantum motors~\cite{fernandez2019}, among others.
On the other hand, the strong Coulomb repulsion between electrons in quantum-dot-based pumps and motors has shown to enhance the performance (or even induce the activation) of these nanodevices~\cite{reckermann2010,calvo2012,calvo2017,ludovico2018,placke2018}. 
The effect of decoherence has also been addressed~\cite{moskalets2001,cremers2002,fernandez2015,fernandez2017} as well as the influence of the friction forces and the system-lead coupling in the dynamics of quantum motors and pumps~\cite{fernandez2017}.
Indeed, the thermodynamics of those systems has proven to be a key aspect to study.
In the last years, different individual efforts have coalesced to give rise to a new field dubbed ``quantum thermodynamics''~\cite{esposito2015,sanchez2016,ludovico2016,ludovico2016B,ludovico2016entropy,benenti2017,whitney2018}, which studies the relations among the different energy fluxes that drive the motion of those machines where quantum mechanics plays a fundamental role.

Despite the progress in the theoretical description of quantum motors and pumps, most of the research has focused on parameter conditions that lie close to the thermodynamic equilibrium, i.e., small bias voltages, temperature gradients, or frequencies of the external driving~\cite{esposito2015,sanchez2016,ludovico2016,ludovico2016B,ludovico2016entropy,benenti2017,whitney2018}.
This is reasonable since under such conditions the linear response regime of the nonequilibrium sources gives an accurate description of the problem, greatly simplifying its general treatment.
For example, in this regime it is common to define dimensionless figures of merit made by some combination of linear response coefficients which give a measurement of the efficiency or the maximum power that quantum devices can achieve.
It is also known that such figures of merit fail in nonlinear regime conditions~\cite{benenti2017}.
Although efforts have been made in this direction, currently there is not a nonlinear version for the figures of merit and the performance must be calculated from the microscopic details of the system's dynamics.
One strategy to deal with such situations is to use phenomenological models where the linear response coefficients are parameterized with respect to the voltage biases, the temperature gradients, or to other relevant parameters of the system, see~\cite{benenti2017} and references therein. However, such parameterizations usually hide the physics behind the nonlinearities and require the optimization of a number of variables that grows very fast with the complexity of the model.
The weakly nonlinear regime of transport has also been explored within the scattering matrix formalism. Under these conditions, it is enough to expand the response coefficients up to second order of the voltage biases and temperature gradients, which can be done by using standard quantum transport techniques. This approach has been applied to a variety of situations, where it proved to be a valuable strategy, see~\cite{sanchez2016} and references therein. However, it would be also important to extend this method to more general situations without hindering the description of the physical processes that take part in the nonlinear effects, while keeping the deep connexions between the response coefficients.

Regarding the dynamics of quantum motors and pumps, one can notice that most of the works in the literature assume a constant terminal velocity of the driving parameters without a concrete model for them.
A typical problem is that when these devices are coupled to nonequilibrium sources, nonconservative current-induced forces (CIFs) appear. These CIFs come, in first place, by assuming a type of Born-Oppenheimer approximation where the electronic and mechanical degrees of freedom can be treated separately, and, secondly, by evaluating the mean value of the force operator~\cite{mozyrsky2006,pistolesi2008,ribeiro2009,dundas2009,suarezmorell2010,lu2010,bennett2010,bode2011,thomas2012,fernandez2015,cunningham2015,lu2015,lu2016,gu2016,bai2016,christensen2016,ludovico2016,ludovico2016B,ludovico2016entropy,fernandez2017,calvo2017,hopjan2018,bruch2018,ludovico2018}. Because of the delayed response of the electronic degrees of freedom to the mechanical motion, one should include the so-called nonadiabatic corrections to the CIFs. This phenomenon is translated into a possible complex dependency of the CIFs on the velocity of the mechanical degrees of freedom. When the effect of the mechanical velocities on CIFs can be treated in linear response, it is clear whether is adequate or not to assume a constant terminal velocity~\cite{fernandez2015,calvo2017,fernandez2017}.
However, in far-from-equilibrium conditions this subject has not been fully addressed.

In this article, we discuss two key aspects of far-from-equilibrium quantum motors and pumps: Their steady-state dynamics, specially when CIFs present nontrivial dependencies on the terminal velocities; and their nonequilibrium thermodynamical properties, when a linear response description is not enough. We provide a systematic expansion to study the 
relations between the different energy fluxes that drive the quantum device. These aspects are illustrated in a concrete example where nonlinear effects due to nonequilibrium sources play a major role on the steady-state properties of the system. We show that these nonlinearities may result in, e.g., negative friction coefficients or motor/pump coexistence regimes.

Our work is organized as follows: In Section~\ref{sec:Langevin} 
we present the general model that describes the considered type of systems and we derive an effective Langevin equation that characterizes the dynamics of the mechanical degrees of freedom, treated classically in the present context.
In Section~\ref{sec:steady_state}, we discuss in general terms the steady-state dynamics of quantum motors and pumps, highlighting some key aspects that differentiate close-to and far-from equilibrium conditions.
In Section~\ref{sec:order-by-order} we derive, on general grounds, the first law of thermodynamics for the kind of systems treated. Then we expand the different energy fluxes passing through the system in terms of the nonequilibrium sources (temperature gradients, bias voltages, and velocities) for arbitrary number of reservoirs.
In this way, we obtain an order-by-order relation between the different energy fluxes entering and leaving the device.
In Section~\ref{sec:efficiency}, we perform a derivation of the rate of entropy production from first principles. Then, based on the second law of thermodynamics, we discuss the limits of the efficiency for different forms of quantum motors and pumps in general nonequilibrium conditions.
In Section~\ref{sec:p-m_rel}, we analyze and give physical interpretation to some of the relations obtained in Section~\ref{sec:order-by-order}.
Finally, in Section~\ref{sec:Coulomb} we consider the CIFs for strongly interacting electrons in a particular example based on a double quantum dot system coupled to a mechanical rotor. We then analyze in detail the effects of higher-order terms in the CIFs on the final steady state of the electromechanical system. 

\section{Current-induced forces and Langevin equation} \label{sec:Langevin}

In this section we introduce the generic model for the treatment of CIFs and the standard method employed
in the description of the dynamics of the mechanical degrees of freedom. As starting point, we consider as \textit{local system} the region where  electronic and mechanical degrees of freedom are present and coupled to each other, like the examples shown in Figure~\ref{fig:1}.
Such a local system is generically modeled by the following Hamiltonian
\begin{equation}
\hat{H}_\tf{local} = \hat{H}_\tf{s}(\hat{\bm{X}})+\frac{\hat{\bm{P}}^{2}}{2m_\tf{eff}}+U(\hat{\bm{X}},t), \label{eq:Hs}
\end{equation}
where $\hat{\bm{X}}=(\hat{X}_{1},...,\hat{X}_{N})$ is the vector of mechanical coordinates and $\hat{\bm{P}}=(\hat{P}_{1},...,\hat{P}_{N})$ collects their associated momenta, $m_\tf{eff}$ is the effective mass related to $\hat{\bm{X}}$, and $U(\hat{\bm{X}},t)$ represents some \textit{external} potential, of mechanical nature, that may be acting on the local system.
The explicit time-dependence on this potential thus emphasizes the fact that an external agent can exert some effective work on the local system.
The Hamiltonian $\hat{H}_\tf{s}$ includes both the electronic degrees of freedom and their coupling to the mechanical ones through
\begin{equation}
\hat{H}_\tf{s}(\hat{\bm{X}}) =  \sum_i E_i(\hat{\bm{X}})\ket{i}\bra{i}, \label{eq:H_s(e)}
\end{equation}
where the sum runs over all possible electronic many-body eigenstates $\ket{i}$.
The local system is then coupled to macroscopic reservoirs and the total Hamiltonian, including the mechanical degrees of freedom, reads
\begin{equation}
\hat{H}_\tf{total} = \hat{H}_\tf{local}+\sum_r \hat{H}_r+\sum_r \hat{H}_{\tf{s},r}. \label{eq:H_total}
\end{equation}
Each lead $r$ is described as a reservoir of noninteracting electrons through the Hamiltonian
\begin{equation}
\hat{H}_r = \sum_{k\sigma} \epsilon_{rk} \hat{c}_{rk\sigma}^\dag \hat{c}_{rk\sigma},
\end{equation}
where $\hat{c}_{rk\sigma}^\dag$ ($\hat{c}_{rk\sigma}$) creates (annihilates) an electron in the $r$-reservoir with state-index $k$ and spin projection $\sigma$.
As usual, the reservoirs are assumed to be always in equilibrium, characterized by a temperature $T_r$ and electrochemical potential $\mu_r$.
The coupling between the local system and the $r$-lead is determined by the tunnel Hamiltonian
\begin{equation}
\hat{H}_{\tf{s},r} = \sum_{k\sigma\ell}  \left( t_{r\ell} \hat{d}_{\ell\sigma}^\dag \hat{c}_{rk\sigma}+\tf{h.c.} \right),
\end{equation}
where $t_{r\ell}$ denotes the tunnel amplitude, assumed to be $k$ and $\sigma$ independent for simplicity, and the fermion operator
$\hat{d}_{\ell\sigma}^\dag$ ($\hat{d}_{\ell\sigma}$) creates (annihilates) one
electron with spin $\sigma$ in the $\ell$-orbital of the local system. The tunnel-coupling strengths $\Gamma_{r\ell} = 2\pi \rho_r |t_{r\ell}|^2$ then characterize the rate at which the electrons enter/leave the local system from/to the $r$-reservoir, where $\rho_r$ is the density of states in the $r$-lead.
Note that $\hat{H}_\tf{s}$ was defined in the eigenstate basis while $\hat{H}_{\tf{s},r}$ is written in terms of single-particle field operators.
The tunnel matrix elements accounting for transitions between different eigenstates can then be obtained from linear superpositions of the above tunnel amplitudes~\cite{leijnse2008}.

To obtain an effective description of the dynamics of the mechanical degrees of freedom through a Langevin equation, we start from the Heisenberg equation of motion for the $\hat{\bm{P}}$ operator, which yields
\begin{equation}
m_\tf{eff}\dt{\dot{\hat{\bm{X}}}}{t}+\hat{\nabla U}(\hat{\bm{X}},t) = -\hat{\nabla H}_\tf{s}(\hat{\bm{X}}). \label{eq:lang_op}
\end{equation}
\new{The measured value $A_\tf{measured}$ of an observable described by an operator $\hat A$ can always be taken as its mean value $A= \braket{\hat{A}}$ plus some fluctuation $\xi_{A}$ around it, i.e., $A_\tf{measured} = A +\xi_{A}$.}
We will work under the nonequilibrium Born-Oppenheimer approximation
~\cite{bennett2010,bode2011,thomas2012,bustos2013,holzbecher2014,arrachea2015,fernandez2015,ludovico2016,fernandez2017}
(or Ehrenfest approximation~\cite{diventra2000,diventra2004,horsfield2004,dundas2009,todorov2010}),
where the dynamics of the electronic and mechanical degrees of freedom
can be separated and the latter is treated classically. This
allows us to neglect the fluctuations of the terms appearing in the left-hand
side of Equation~(\ref{eq:lang_op}), and describe the mechanical motion only through the mean value $\bm{X}$, which is reasonable for large or massive objects. With this in mind, we obtain the following Langevin equation of motion:
\begin{equation}
m_\tf{eff}\dt{\dot{\bm{X}}}{t}+\bm{F}_\tf{ext} = \bm{F}+\bm{\xi}, \label{eq:lang}
\end{equation}
where $\bm{F} = -\left\langle \hat{\nabla H}_\tf{s}\right\rangle = i \left\langle [\hat{H}_\tf{s}(\hat{\bm{X}}),\hat{\bm{P}}]\right\rangle $
and $\bm{\xi}$ account for the mean value and the fluctuation of
the CIF, respectively.\footnote{Throughout this manuscript we take $\hbar=1$ for simplicity.}
As we shall see later on, the external force applied to the mechanical part of the local system, $\bm{F}_\tf{ext}$, plays the role of an eventual ``load'' force for a quantum motor or a ``driving'' force for a quantum pump.
As this force will be typically opposed to the CIF, we define $\bm{F}_\tf{ext}$ with a minus sign for better clarity in future discussions.
The main task, therefore, relies on the calculation of the CIFs from appropriate formalisms capable to describe the dynamics of the electronic part of the system.
Once these forces are calculated, we can use Equation~(\ref{eq:lang}) to integrate the classical equations of motion and 
obtain the effective dynamics of the complete electromechanical system.

In most previous works $\bm{F}$ is expanded up to first order in $\dot{\bm{X}}$, i.e., $\bm{F} \approx 
\bm{F}^{(0)} - \bm{\gamma} \cdot \dot{\bm{X}}$.
The resulting CIF is then the sum of an adiabatic contribution $\bm{F}^{(0)}$ and its first nonadiabatic correction
$\bm{F}^{(1)} = -\boldsymbol{\gamma}\dot{\bm{X}}$, respectively. Under this approximation, Equation~(\ref{eq:lang}) turns into
\begin{equation}
m_\tf{eff}\dt{\dot{\bm{X}}}{t}+\bm{F}_\tf{ext} = \bm{F}^{(0)}-\bm{\gamma}\cdot\dot{\bm{X}}+\bm{\xi}. \label{eq:lang_1st}
\end{equation}
Explicit formulas for the calculation of $\bm{F}^{(0)}$, $\bm{\gamma}$, and $\bm{\xi}$
in terms of Green functions and scattering matrices were derived
in~\cite{mozyrsky2006,pistolesi2008,bennett2010,bode2011,thomas2012}, and extended in~\cite{fernandez2015,fernandez2017} to account for decoherent events.
Although these expressions were obtained in the context of noninteracting particles, they can be used in effective Hamiltonians derived from first principles calculations~\cite{ribeiro2009,suarezmorell2010}.
In~\cite{ludovico2016,ludovico2016B,ludovico2016entropy} the CIFs were obtained from the Floquet Green's function formalism.
The role of Coulomb interactions was addressed through different formalisms and methods like, e.g., many-body perturbation theory based on nonequilibrium Green's functions~\cite{hopjan2018}; modeling the system as a Luttinger liquid~\cite{bruch2018}; and using a time-dependent slave-boson approximation~\cite{ludovico2018}.
In~\cite{calvo2017}, explicit expressions for the CIFs within the Coulomb blockade regime of transport were obtained using a real-time diagrammatic approach~\cite{splettstoesser2006}, which we present in more detail in Section~\ref{sec:Coulomb} when considering the example of Figure~\ref{fig:1}c.

\section{Mechanical steady-state } \label{sec:steady_state}

In this paper, we will restrict ourselves to systems that perform overall cyclic motions.
Immediate examples are shown in Figure~\ref{fig:1}a and c, where the rotation angle of the rotor can be assigned as the natural mechanical coordinate. 
On the other hand, the examples shown in Figure~\ref{fig:1}b and d may also, under certain circumstances, sustain cyclic motion, though the general coordinate could be not so obvious.  
As a possibility for the quantum shuttle of Figure~\ref{fig:1}b, the cyclic motion would involve a cyclic reversal of the bias voltage (ac-driven). This ac-driven case, though intriguing, goes beyond the scope of the present manuscript as we are not considering here time-dependent biases.
Another scenario would be that of Figure~\ref{fig:1}d, where the cyclic motion is in principle attainable by periodically hitting the device.
\new{Note that we are not dealing with the steady-state of sets of interacting nanomotors, such as those described in, e.g., Refs.~\cite{antoni1995,levin2014,atenas2017}. Instead, here we are interested in the steady-state of the mechanical part of isolated quantum motors and pumps that interact solely with the electrons of a set of reservoirs, and where, at most, Coulomb interactions are only taken into account within the local system.}

To discuss the dynamics of cyclic motions in simple terms, we start by projecting Equation~(\ref{eq:lang})
on a closed trajectory defined in the space of $\bm{X}$.
By assuming a circular trajectory, the dynamics can be described by an angle $\theta$, its associated angular velocity $\dot{\theta}$, the moment of inertia $\tc{I}$, and the torques
$\tc{F}$, $\tc{F}_\tf{ext}$, and $\xi_{\theta}$.
Using this, we obtain an effective angular Langevin equation equivalent to Equation~(\ref{eq:lang}),
\begin{equation}
\dt{\dot\theta}{t} = \frac{1}{\tc{I}} \left[ \tc{F} - \tc{F}_\tf{ext} +\xi_\theta \right] . \label{eq:lang_ang}
\end{equation}
We assume that after a long waiting time the system arrives to the steady-state regime where the mechanical motion becomes periodic, and it is then characterized by a time period $\tau$ such that $\theta(t+\tau)=\theta(t)$ and $\dot{\theta}(t+\tau)=\dot{\theta}(t)$.
Moreover, we will assume that the stochastic force plays a minor role in the above equation, such that it does not affect the mean values of the dynamical variables $\theta$ and $\dot\theta$, i.e., the mean trajectories with or without the stochastic force approximately coincide.
This occurs, for example, at low temperatures or in mechanical systems with a large moment of inertia~\cite{fernandez2015,fernandez2017,calvo2017}.
In the following we will just ignore $\xi_\theta$ for practical purposes.\footnote{
This is the opposite regime of another type of nanomotors: The Brownian motors~\cite{martinez2016,spiechowicz2014}.
}
Under the above assumptions, we integrate both sides of Equation~(\ref{eq:lang_ang}) from an initial position $\theta_i$ to a final one $\theta_f$ and obtain:
\begin{eqnarray}
\frac{\tc{I}}{2}\left[\dot{\theta}_f^2-\dot{\theta}_i^2\right] & = & \int_{\theta_i}^{\theta_f}
\left [ \tc{F}  - \tc{F}_\tf{ext} \right ] \tf{d}\theta. \label{eq:DeltaK}
\end{eqnarray}
The torques in this equation are, in general, intricate functions of both $\theta$ and $\dot{\theta}$.
Therefore, the calculation of the $\theta$-dependent angular velocity usually requires the resolution of a transcendental equation (see, e.g.,~\cite{calvo2017}).
Alternatively, one can obtain $\theta(t)$ from the numerical integration of the equation of motion by standard techniques like, e.g., the Runge-Kutta method.
All this greatly complicates the study of quantum motors and pumps, to the point where it becomes almost impossible to draw any general conclusion.
For this reason, one common simplification consists in taking the terminal velocity as constant during the whole cycle
~\cite{bustos2013,arrachea2015,fernandez2015,ludovico2016entropy,ludovico2016,ludovico2016B,fernandez2017,bruch2018,ludovico2018,fernandez2019}.
Indeed, this description is exact if the external agent \textit{compels} the constant velocity condition to be fulfilled in a controllable manner, as is often conceived in quantum pumping protocols. However, this is not the case in general, and typically one expects internal variations for $\dot\theta$ in one period. We now address this interesting issue in more detail. First, we take the integral in Equation~(\ref{eq:DeltaK}) over the whole period. This gives
\begin{equation}
 \tc{W}_\tf{ext} = \tc{W}_F, \quad \tf{where} \quad \tc{W}_{F}= \int_0^\tau \tc{F} \, \dot\theta \, \tf{d}t, \quad \tf{and} \quad
 \tc{W}_\tf{ext} = \int_0^\tau \tc{F}_\tf{ext} \, \dot\theta \, \tf{d}t.
\label{eq:s-s_condition}
\end{equation}
The above stationary state condition thus establishes that the work originated from the CIF is always compensated by the \new{external} mechanical work in the case this regime can be reached.
Now, let us assume for a moment that the terminal velocity of a nanodevice is constant and positive (we leave the discussion of the effect of the sign of $\dot\theta$ for later when treating a concrete example in Section~\ref{sec:DQD_example}).
If we now expand $\tc{F}$ in terms of $\dot{\theta}$, Equation~(\ref{eq:s-s_condition}) yields
\begin{eqnarray}
\mathcal{W}_{\tf{ext}} & = & \sum_{k}\left(\int_{0}^{2\pi}\left. \dd{^k \tc{F}}{\dot\theta^k}\right|_{\dot{\theta}=0} 
\frac{\tf{d}\theta}{k!} \right) \dot{\theta}^{k} . \label{eq:const_vel_condit}
\end{eqnarray}
Two important conclusions can be extracted from the above formal solution.
First, there may be conditions where some roots of Equation~(\ref{eq:const_vel_condit}) are complex numbers, meaning that
the assumption $\dot{\theta} = \tf{const.}$ is nonsense, as the periodicity condition required for the steady-state regime would not be fulfilled. 
Second, for real solutions, it was shown in~\cite{calvo2017,fernandez2017} that the moment of inertia $\tc{I}$ not only affects the time that takes to the mechanical system to reach the stationary regime, but also the internal range in the angular velocity, i.e., the difference $\Delta \dot\theta = \dot\theta_\tf{max} - \dot\theta_\tf{min}$ in one period.
According to Equation~(\ref{eq:const_vel_condit}), $\dot{\theta}$ is independent of $\tc{I}$, while the variation of 
$\dot{\theta}$ scales with $\tc{I}^{-1}$, cf. Equation~(\ref{eq:DeltaK}).
Then the ratio $\Delta \dot{\theta}/\dot\theta$, which is the relevant quantity in our analysis, should vanish for large $\tc{I}$ values, justifying the constant velocity assumption for large or massive mechanical systems.

The value of $\tc{F}_{\tf{ext}}$ is supposed to be controllable externally as well as the voltage and temperature biases which, in turn, affect the current-induced torque $\tc{F}$.
Therefore, under the above discussed conditions, $\dot{\theta}$ can be thought as a parameter that surely depends on the internal details of the system, but it is also tunable by external ``knobs''. Let us analyze a concrete example:
Consider a local system connected to two leads at the same temperature and with a small bias voltage $eV=\mu_L-\mu_R$. By considering the current induced torque up to its first nonadiabatic correction, i.e., $\tc{F} \approx \tc{F}^{(0)} - \gamma \dot{\theta}$,
and assuming that $\tc{F}_\tf{ext}$ is independent of $\dot\theta$, the following relation must hold, according to the above discussion,
\begin{equation}
\dot\theta \approx \frac{Q_{I_R}}{2\pi \bar\gamma} \left( V- \frac{\tc{W}_\tf{ext}}{Q_{I_R}} \right) = \frac{Q_{I_R} 
V_\tf{eff}}{2\pi \bar\gamma}, \label{eq_tauWconst}
\end{equation}
where $\bar\gamma$ is the average electronic friction coefficient along the cycle, $Q_{I_R}$ is the pumped charge to the right lead, and we used the Onsager's reciprocal relation between $\tc{F}$ and the charge current $I_R$ in the absence of magnetic fields~\cite{fernandez2015,ludovico2016,calvo2017,fernandez2017}.
Alternatively, if we assume that the \new{external} torque is of the form $\tc{F}_\tf{ext} = \gamma_\tf{ext}\dot\theta$, one finds
\begin{equation}
\dot{\theta} \approx \frac{Q_{I_R} V}{2\pi} \frac{1}{(\bar{\gamma}+\bar{\gamma}_\tf{ext})} = \frac{Q_{I_R} V}{2\pi \gamma_\tf{eff}}.\label{eq_tauWfricc}
\end{equation}
Note in the above equations that, at least in the present order, the effect of the \new{external} forces can be described as a renormalization of the bias voltage $V$ or the electronic friction coefficient $\gamma$.
Numerical simulations in~\cite{fernandez2017,calvo2017} show that above equations agree well in general with the steady-state velocities found by integrating the equation of motion.
However, at very small voltages, essential differences may appear.
There is a critical voltage below which the dissipated energy per cycle can not be compensated by the work done by the CIF and thus $\dot\theta = 0$. We dubbed this the ``nonoperational'' regime of the motor.
Moreover, when increasing the bias voltage, there is an intermediate region where a hysteresis cycle appears, and two values of the velocity are possible ($\dot\theta = 0$ and those given
by the above equations).
Although in Section~\ref{sec:DQD_example}, we will take $\dot\theta$ as constant when discussing a specific example, the reader should keep in mind that this approximation not always holds, especially at very small voltages or $\tc{I}$.

\section{Order-by-order energy conservation} \label{sec:order-by-order}

In the previous sections we introduced and discussed the role of the mechanical degrees of freedom, emphasizing certain parameter restrictions which allow for the simplification in their dynamics. In this section we are going to derive, on general grounds, essential relations between the electronic and mechanical degrees of freedom from the point of view of energy conservation. Importantly, we will focus in sistematic expansions beyond the standard linear regime of nonequilibrium sources like, e.g., the mechanical velocity, the bias voltage and the temperature gradient in systems composed by an arbitrary number of reservoirs.

As already pointed out, we are treating the mechanical degrees of freedom classically, such that their effect on the electronic degrees of freedom enters as a parametric dependence in the electronic part of the total Hamiltonian, which now reads
\begin{equation}
\hat{H} = \sum_r \hat{H}_r + \hat{H}_\tf{s} + \sum_r \hat{H}_{\tf{s},r}.
\end{equation}
The mechanical part of the local system, when treated classically in Equation~(\ref{eq:lang}), introduces an explicit time dependence into $\hat{H}_\tf{s}$ that, in turn, it makes $\tf{d}\braket{\hat{H}}/\tf{d}t \neq 0$. According to the above Hamiltonian, the total internal energy of the electronic system, $U=\braket{\hat{H}}$, can be split into energy contributions from the reservoirs, the local system, and the tunnel couplings.
The time-variation of the internal energies associated with the different partitions of the system are:
\begin{equation}
\dot{U}_\beta	= J_\beta^E + \bigg\langle \dd{\hat{H}_\beta}{t} \bigg\rangle, \quad \tf{where} \quad 
J_\beta^E = i \braket{\left[\hat{H},\hat{H}_\beta \right]}
\label{eq:dotU_l}
\end{equation}
is the mean value of the energy flux entering in the subsystem $\beta= \{r , (\tf{s},r), \tf{s} \}$.
The value of $\braket{\partial_t \hat{H}_\beta}$ is zero when we evaluate the energy flux in the reservoirs and the tunnel couplings, but equals 
$-\dot{\tc{W}}_F$ when $\beta$ is evaluated in the local system.\footnote{Although strictly speaking $\tc{W}_F$ is time-independent when considering a closed trajectory, we use the symbol $\dot{\tc{W}}_F$ to denote the power delivered by the CIF, i.e., $\dot{\tc{W}}_F = - \braket{\partial_t \hat{H}_\tf{s}} = \bm{F} \cdot \dot{\bm{X}}$.} 
The latter comes from the definition of the CIF given in Equation~(\ref{eq:lang}).
Note that the energy fluxes fulfill the condition $\sum_r (J_r^E+J_{\tf{s},r}^E)+J_\tf{s}^E = 0$. Therefore, the variation of the total internal energy of the electrons yields the following conservation rule
\begin{equation}
\dot{U} = \sum_r \dot{U}_r+\sum_r \dot{U}_{\tf{s},r}+\dot{U}_\tf{s}  = \sum_r \left( J_r^E + J_{\tf{s},r}^E \right) + J_\tf{s}^E -\dot{\tc{W}}_F = -\dot{\tc{W}}_F. \label{eq:E_conservation}
\end{equation}
Conservation of the total number of particles implies a relation between the particle currents of the reservoirs $\dot{N}_r$ and that of the local system 
$\dot{N}_\tf{s}$
\begin{equation}
\dot{N}_\tf{total} = \dt{}{t} \braket{\hat{N}_\tf{total}} = \sum_r \dt{}{t} \braket{\hat{N}_r} + 
\dt{}{t} \braket{\hat{N}_\tf{s}} = 0 \quad \Rightarrow \quad \sum_r \dot{N}_r = -\dot{N}_\tf{s}. \label{eq:N_conservation}
\end{equation}
This is so since no particle can be assigned to the coupling region.
Now, let us assume the system is in the steady-state regime and integrate Equation~(\ref{eq:E_conservation}) 
over a time period $\tau$ of the cyclic motion. Under this condition, the above defined local quantities only depend periodically on time, i.e., $U_\tf{s}(t+\tau) = U_\tf{s}(t)$, $U_{\tf{s},r}(t+\tau) = U_{\tf{s},r}(t)$, and $\dot{N}_\tf{s}(t+\tau) = \dot{N}_\tf{s}(t)$. Therefore, the following quantities should evaluate to zero, i.e.,
\begin{equation}
\int_0^\tau\dot{U}_{\tf{s},r} \, \tf{d}t = 0 , \quad 
\int_0^\tau\dot{U}_\tf{s} \, \tf{d}t = 0, \quad \tf{and} \quad
\int_0^\tau\dot{N}_\tf{s} \, \tf{d}t = 0,  \label{eq:dotU}
\end{equation}
as these are integrals of a total derivative of some periodic function. This means that no energy (or particles) is accumulated/extracted indefinitely within the finite regions defined by the local system or its coupling to the leads. 
Equation~(\ref{eq:dotU}) can be used together with Equation~(\ref{eq:N_conservation})
to prove charge current conservation between reservoirs, $\sum_{r} \int_0^\tau I_{r} \, \tf{d}t = 0$, where the charge current of the $r$-reservoir is defined as $I_r = e \dot{N}_r$, with $e>0$ being minus the electron's charge.

We will take the following definition for the heat current $J_r$ in the reservoir $r$,\footnote{
In \cite{ludovico2014,ludovico2016B,ludovico2018B} the authors proposed a different definition for the heat current of the reservoirs, $J_{r}=J_{r}^{E}+\left(J_{\tf{s},r}^{E}/2\right)-\mu_{r}\dot{N}_{r}$.
However, the inclusion or not of half the heat current of the coupling region, $\left(J_{\tf{s},r}^{E}/2\right)$, does not make any difference in the present paper as this quantity integrates to zero over a cycle, Equation~(\ref{eq:dotU}), and does not contribute to the rate of entropy production as we will see in the next section, Equation~(\ref{eq:S_rev}).
}
\begin{equation}
J_r = J_r^E - \mu_r \dot{N}_r. \label{eq:J_definition}
\end{equation}
Replacing this expression in Equation~(\ref{eq:E_conservation}), and integrating over a period, results in
\begin{equation}
\sum_r \left(Q_{J_r}+Q_{I_r}\delta V_r\right) = -\mathcal{W}_{F}, \quad \tf{where} \quad Q_{J_r}=\int_0^\tau J_r \, \tf{d}t, 
\quad \tf{and} \quad Q_{I_r}=\int_0^\tau I_r \, \tf{d}t,
\label{eq:1st_Law}
\end{equation}
where $\delta V_r = (\mu_r - \mu_0)/e$ and $\mu_0$ is an arbitrary reference's potential.
We, in addition, defined the quantities $Q_{J_r}$ and $Q_{I_r}$ which are, respectively, the total heat and charge pumped to reservoir $r$ in a cycle.
Note that in the above equation we have used conservation of the total charge.

Before to continue, we would like to emphasize that the periodic motion imposed by the steady-state regime can be further exploited to reduce the number of mechanical coordinates to an effective description of the CIF. In principle, one can always recognize a generalized coordinate $\chi$, that parameterizes the closed mechanical trajectory. This, for example, can be accounted for by taking $\chi = \Omega \, \tf{mod}(t,\tau)$, such that $\dot\chi = \Omega = 2\pi/\tau$. In other words, one can always find a natural scale, $\Omega$ in the present case, for all mechanical velocities along the trajectory. We will take this natural scale as the expansion variable for the CIF.
The current-induced work then reads $\tc{W}_F = \int_0^\tau F_\Omega \, \Omega \, \tf{d}t$, where 
$F_\Omega = \bm{F} \cdot \partial_\chi \bm{X}$. Of course, with this $\chi$-parameterization, the velocity is constant by definition, but at the expense that now the calculation of $F_\Omega$ requires the knowledge of the mechanical trajectory.
\footnote{
For circular trajectories and the conditions discussed in Section~\ref{sec:steady_state}, the coordinate $\chi$ would be 
$|\theta|$. For other cases, it could not be that simple to find $\chi$ and one should first solve the system's equation of motion.
}

In addition to the mechanical velocity $\Omega$, the CIF and the currents can also be expanded in terms of the remaining nonequilibrium sources, 
corresponding to voltage and temperature deviations $\delta \bm{V} = (\delta V_1, \delta V_2,...)^\tf{T}$ and $\delta \bm{T} = (\delta T_1, \delta T_2,...)^\tf{T}$, respectively, from their equilibrium values $V_\tf{eq}$ and $T_\tf{eq}$. For an arbitrary observable $R$, this general expansion takes the form:
\begin{equation}
R = \sum_{|\bm{\alpha}| \geq 0} R^{\bm{\alpha}} = \sum_{|\bm{\alpha}| \geq 0} \frac{(\Omega,\delta \bm V, \delta \bm{T})^{\bm{\alpha}}}{\bm{\alpha}!}
\left( \partial^{\bm{\alpha}} R \right)_\tf{eq}, \label{eq:expansion}
\end{equation}
where we used the following multi-index notation: $\bm{\alpha}=\left(n_{\Omega},n_{V_1},...,n_{T_1},...\right)$;
$\bm{\alpha}!=n_{\Omega}!n_{V_1}!...n_{T_1}!...$; $(\Omega,\delta\bm{V},\delta\bm{T})^{\bm{\alpha}} = \Omega^{n_\Omega}\delta V_1^{n_{V_1}}...\delta T_1^{n_{T_1}}...$;
and
\begin{equation}
\partial^{\bm{\alpha}} = \frac{\partial^{\left(n_\Omega+n_{V_1}+...n_{T_1}+...\right)}}{{\partial\Omega}^{n_\Omega}\partial V_1^{n_{V_1}}...\partial T_1^{n_{T_1}}...}.
\end{equation}
This, in turn, allows us to recognize in the integral quantities of Equation~(\ref{eq:1st_Law}) the following expansion coefficients
\begin{equation}
\tc{W}_F^{\bm{\alpha}} = \int_0^\tau F_\Omega^{\bm{\alpha}} \Omega \tf{d}t, \quad Q_{J_r}^{\bm{\alpha}} = \int_0^\tau J_r^{\bm{\alpha}} \tf{d}t,
\quad \tf{and} \quad Q_{I_r}^{\bm{\alpha}} = \int_0^\tau I_r^{\bm{\alpha}} \tf{d}t, \label{eq:int_exp}
\end{equation}
where $F_\Omega^{\bm{\alpha}}$, $J_r^{\bm{\alpha}}$, and $I_r^{\bm{\alpha}}$ all have the form given by Equation~(\ref{eq:expansion}). To find a consistent order-by-order conservation relation from Equation~(\ref{eq:1st_Law}), we should first note that the involved terms enter in different 
orders: $\tf{W}_F^{\bm{\alpha}}$ contains an additional $\Omega$ from the time-integral as compared with the pumped currents, while the term $Q_{I_r}^{\bm{\alpha}}\delta V_r$ is one order higher in $\delta V_r$ than the other two. Therefore, the following relation must hold for every order 
$\bm{\alpha}$ of the expansion
\begin{equation}
\bm{Q}_J^{\bm{\alpha}} \cdot \bm{1} + \bm{Q}_I^{\bm{\alpha}(\bm{V})} \cdot \delta\bm{V} = - \tc{W}_F^{\bm{\alpha}(\Omega)}, \label{eq:1stLaw_vect}
\end{equation}
where $\bm{1}=(1,1,...)^\tf{T}$, $\bm{Q}_J^{\bm{\alpha}}= (Q_{J_1}^{\bm{\alpha}},Q_{J_2}^{\bm{\alpha}},...)$,
$\bm{Q}_I^{\bm{\alpha}(\bm{V})} = (Q_{I_1}^{\bm{\alpha}(V_1)},Q_{I_2}^{\bm{\alpha}(V_2)},...)$,
and we used the shorthand $\bm{\alpha}(\Omega) = (n_\Omega-1,n_{V_1},...,n_{T_1},...)$
and $\bm{\alpha}(V_r) = (n_\Omega,n_{V_1},...,n_{V_r}-1,...,n_{T_1},...)$. Equation~(\ref{eq:1stLaw_vect}) results important for a systematic study of far-from-equilibrium systems as it helps to rationalize how every order of the expansion of an energy flux is connected with the others. The equation is completely general and valid for any value of the relevant quantities $\Omega$, $\delta \bm{V}$, and $\delta \bm{T}$. Importantly, the above conservation rule can be extended to other nonequilibrium sources like, e.g., spin polarization in ferromagnetic leads, such that other type of currents could also be considered in the energy transfer process between electronic and mechanical parts of the system.

\section{Entropy production and efficiency} \label{sec:efficiency}

In 1865, Rudolf Clausius proposed a new state function, the thermodynamic entropy $S$, that turned out to be crucial to study the limits and the efficiency of different physical processes.
The thermodynamic entropy is defined as the amount of heat $\delta Q_J$ which is transferred in a reversible thermodynamic process, 
$\delta S = \delta Q_\tf{rev}/T$.
Here we are assuming that each reservoir $r$ is in its own equilibrium at a constant temperature $T_r$ and chemical potential $\mu_r$ (we are not going to treat time-dependent $T_r$ or $\mu_r$). As the reservoirs are considered to be macroscopic, the aforementioned equilibrium state is not altered by the coupling to the local system. This assumption allows us to associate the reservoirs' heat flux with the variation in their thermodynamic entropy, i.e., $\delta Q_{J_r} = T_r \delta S_r$.
\footnote{
The information theory entropy, known as the Shannon or von Neumann entropy, times the Boltzmann constant equals the thermodynamic entropy for equilibrium states.
There is a debate on whether this equality can be extended to nonequilibrium states, see, e.g.~\cite{plenio2001,deutsch2010,gemmer2014}.
However, for the purpose of this article, only the change of the entropy of the reservoirs is needed, not that of the system.
Besides, we will not need to evaluate the entropy from the density matrix.
Therefore, the thermodynamic definition of entropy suffices in our case.
}
Therefore, from Equation~(\ref{eq:dotU_l}) and the definition of the heat current given in Equation~(\ref{eq:J_definition}), we can write
\begin{equation}
\dot{U}_r = T_r \dot S_r+ \dot{N}_r \mu_r.
\end{equation}
Now, let us consider the sum of the internal energy over the set of all reservoirs coupled to the local system,
\begin{equation}
\sum_r \dot{U}_r = \sum_ r T_r \dot{S}_r + \sum_r \dot{N}_r \mu_r.
\end{equation}
If we take $T_{r}=\delta T_{r}+T_{0}$ and $\mu_{r}=\mu_{0}+\delta\mu_{r}$,
\footnote{The values of $\mu_{0}$ and $T_{0}$ are completely arbitrary, and there is no need to identify them with the chemical potential and temperature of the central region, which can be ill-defined far from
equilibrium.} add and subtract the change of the internal energy of the local system $\dot{U}_{s}$ and that of the couplings between the local system and the reservoirs $\sum_{r}\dot{U}_{\tf{s},r}$, one can rearrange the above equation to the following:
\begin{equation}
T_0 \sum_r \dot{S}_r = \left( \dot{U}_\tf{s}+\sum_r \dot{U}_r+\sum_r \dot{U}_{\tf{s},r} \right) - \sum_r \dot{N}_r \delta\mu_r-\sum_r \dot{S}_r
\delta T_r - \left(\dot{U}_\tf{s}+\sum_r \dot{U}_{\tf{s},r} \right) - \mu_0 \left(\sum_r \dot{N}_r \right).\label{eq:dotS_01}
\end{equation}
Replacing $\dot{S}_{r}$ by $=J_{r}/T_{r}$ in the right-hand side of the above equation, using energy
conservation (\ref{eq:E_conservation}), and particle number conservation
(\ref{eq:N_conservation}), allows one to rewrite Equation~(\ref{eq:dotS_01})
as
\begin{eqnarray}
T_{0}\dot{S}_{\tf{res}} & = & -F_\Omega \Omega-\sum_{r}I_{r}\delta V_{r}-\sum_{r}J_{r}\left(\frac{\delta T_{r}}{T_{r}}\right)-\left(\dot{U}_\tf{s}+\sum_{r}\dot{U}_{\tf{s},r}\right)+\mu_{0}\dot{N}_\tf{s},
\end{eqnarray}
where $\dot{S}_{\tf{res}} = \sum_{r}\dot{S}_{r}$ is the variation of the entropy of the electrons of all reservoirs, and we used $e\dot{N}_{r}=I_{r}$.
The CIF can be split into ``equilibrium'' and ``nonequilibrium'' terms, $F^{\tf{(eq)}}$ and $F^{\tf{(ne)}}$ respectively, where one can prove that $F^{\tf{(eq)}}$
is always conservative~\cite{dundas2009,bustos2013,fernandez2017,calvo2017}.
We are interested in the steady-state situation of our local system. As discussed around Equation~(\ref{eq:dotU}), the change of the internal energy of the electronic part of the local system and that of the coupling region must be zero after a cycle, as energy cannot be accumulated indefinitely within a finite region.
The same argument is true for the number of particles accumulated in a cycle, which should be zero. At steady state, we therefore recognize the reversible component of the entropy variation as that given by
\begin{equation}
\dot{S}_{\tf{res}}^{\left(\tf{rev}\right)} =-\frac{1}{T_0} \left(F^{\tf{(eq)}}_\Omega \Omega+\dot{U}_\tf{s}+\sum_{r}\dot{U}_{\tf{s},r}-\mu_{0}\dot{N}_\tf{s}\right).\label{eq:S_rev}
\end{equation}
Obviously, this quantity will not contribute to the total entropy production. Therefore, the rate of entropy production $\dot{S}_{\tf{res}}^{\left(\tf{irrev}\right)}$ yields
\begin{equation}
\dot{S}_\tf{res}^{(\tf{irrev})} = - \frac{1}{T_0} \left( F^\tf{(ne)}_\Omega \Omega - \bm{I} \cdot \delta\bm{V}-\bm{J} \cdot \delta\bm{\tc{T}} \right), \label{eq:dotS_general}
\end{equation}
where $\delta \bm{\tc{T}} = (\delta T_1 / T_1, \delta T_2/T_2,...)^\tf{T}$, and the currents are defined through $\bm{I} = (I_1,I_2,...)$, and $\bm{J} = (J_1,J_2,...)$. Integrating Equation~(\ref{eq:dotS_general}) over a cycle and taking into account the second law of thermodynamics, one finds
\begin{equation}
0 \geq \tc{W}_F + \bm{Q}_I \cdot \delta \bm{V} + \bm{Q}_J \cdot \delta \bm{\tc{T}}. \label{eq:2nd_law}
\end{equation}
The above general formula, also valid far from equilibrium, can be used to set efficiency bounds for energy transfer processes between the electronic and mechanical degrees of freedom.
For example, if we take $\delta\bm{\tc{T}} = 0$, $\bm{Q}_I \cdot \delta \bm{V} < 0$, and $\tc{W}_F>0$, then the system should operate as a nanomotor driven by electric currents, and the following relation holds
\begin{equation}
1 \geq -\frac{\tc{W}_F}{\bm{Q}_I \cdot \delta \bm{V}}, \label{eq:electric-motor}
\end{equation}
while for $\bm{Q}_I \cdot \delta\bm{V}>0$ and $\tc{W}_F<0$, the system operates as a charge pump, and Equation~(\ref{eq:2nd_law}) implies
\begin{equation}
1 \geq -\frac{\bm{Q}_I \cdot \delta \bm{V}}{\tc{W}_F}.
\end{equation}
Notice that, because of the steady-state condition, $\tc{W}_F$ equals $\tc{W}_\tf{ext}$, where $\tc{W}_\tf{ext}$ can be taken as the output or the input energy, depending on the considered type of process.
Therefore, the above formulas describe the efficiency $\eta$ of the device's process, defined as the ratio between the output and input energies per cycle. It is also interesting to note that the above equations reflect no more than energy conservation in this particular case.
A different situation occurs for $\delta\bm{V}=0$ and $\delta\bm{\tc{T}} \neq 0$, where Equation~(\ref{eq:2nd_law}) yields
\begin{equation}
1 \geq -\frac{\tc{W}_F}{\bm{Q}_J \cdot \delta\bm{\tc{T}}}, \quad \tf{and} \quad
1 \geq -\frac{\bm{Q}_J \cdot \delta\bm{\tc{T}}}{\tc{W}_F}. \label{eq:heat-motor}
\end{equation}
The first equation thus corresponds to a quantum heat engine and the second one to a quantum heat pump, respectively. Now, because of the factor $\delta\bm{\tc{T}}$, the above formulas differ from what it is expected from energy conservation solely. This is clear in a two-lead system, where 
$\eta$ is limited by the Carnot's efficiency of heat engines and refrigerators, respectively. To illustrate this, let us consider a hot and a cold reservoir, and set the temperature of the cold reservoir as the reference. For the heat engine, this gives
\begin{equation}
\bm{Q}_J \cdot \delta\bm{\tc{T}} = Q_{J_\tf{hot}} \left(1-\frac{T_\tf{cold}}{T_\tf{hot}}\right) < 0 \quad \Rightarrow \quad 
\left(1-\frac{T_\tf{cold}}{T_\tf{hot}}\right) \geq 
-\frac{\tc{W}_F}{Q_{J_\tf{hot}}},
\end{equation}
where the left-hand side of the second equation represents the Carnot limit for heat engines.
Other energy transfer processes mixing voltage and temperature biases can also be analyzed in the context of Equation~(\ref{eq:2nd_law}) to set the bounds of their associated efficiencies.

\section{Pump-motor relations} \label{sec:p-m_rel}

It is clear from Equation~(\ref{eq:1stLaw_vect}) that there is an infinite number of relations that can be used to connect the pumped heat or charge and the work done by the CIF.
In this section we give some physical interpretation to the leading orders in the general expansion, which highlights the utility of Equation~(\ref{eq:1stLaw_vect}).

We start from the order-by-order relations by taking $|\bm{\alpha}|=0$. Importantly, Equation~(\ref{eq:1stLaw_vect}) with $|\bm{\alpha}|=0$ seems to impose the evaluation of terms with negative coefficients. For example, in the exponent of $\bm{Q}_I$ one would be tempted to evaluate $\bm{\alpha}(V_1)=(0,-1,0,0...)$, but this term is not defined in the expansions of Equation~(\ref{eq:int_exp}), and then it should be taken as zero. Therefore, 
Equation~(\ref{eq:1stLaw_vect}) yields,
\begin{equation}
\bm{Q}_J^{\bm{0}} \cdot \bm{1} = \sum_r \int_0^\tau \left( J_r \right)_\tf{eq} \, \tf{d}t = 0. \label{eq:alpha0}
\end{equation}
This simply reflects the fact that no net heat current occurs at equilibrium.
For $|\bm{\alpha}|=1$ all relations are summarized in the following three cases:
\begin{equation}
\sum_r \int_0^\tau \left(\dd{J_r}{\Omega}\right)_\tf{eq} \Omega \, \tf{d}t = 0, \quad
\sum_r \int_0^\tau \left(\dd{J_r}{V_i}   \right)_\tf{eq} \delta V_i \, \tf{d}t = 0, \quad \tf{and} \quad
\sum_r \int_0^\tau \left(\dd{J_r}{T_i}   \right)_\tf{eq} \delta T_i \, \tf{d}t = 0,
\label{eq:alpha1_case1}
\end{equation}
where we used the fact that equilibrium forces are conservative~\cite{dundas2009,bustos2013,ludovico2016B,calvo2017,fernandez2017},
i.e, $\int_0^\tau F_\Omega^\tf{(eq)} \Omega \tf{d}t=0$,
and that there are no net charge currents in equilibrium, thus $\int_0^\tau (I_r)_\tf{eq} \tf{d}t=0$. The above relations mean that, at first order, there is a conservation of pumped heat between reservoirs.
For $|\bm{\alpha}|=2$, there are many relations and cases, but we restrict ourselves to only a few of them. For $n_\Omega=2$, 
$n_{V_i}=0$, and $n_{T_j}=0$, Equation~(\ref{eq:1stLaw_vect}) gives
\begin{eqnarray}
\sum_{r}\intop_{0}^{\tau}\frac{{\Omega}^{2}}{2}\left(\frac{\partial^{2}J_{r}}{\partial{\Omega}^{2}}\right)_{\tf{eq}}\tf{d}t & = & -\intop_{0}^{\tau}\left(\frac{\partial F_{\Omega}}{\partial{\Omega}}\right)_{\tf{eq}}{\Omega}^{2}\tf{d}t.
\label{eq:alpha2_case1}
\end{eqnarray}
Now, the quantity $(\partial_\Omega F_\Omega)_\tf{eq}$ is minus the electronic friction coefficient at equilibrium. Therefore, this relation shows 
that the energy dissipated as friction in the motor is delivered as heat to the reservoirs. More precisely, as second order pumped heat. For $n_\Omega=0$, $n_{V_i}=1$, $n_{V_j}=1$, and $n_{T_k}=0$, Equation~(\ref{eq:1stLaw_vect}) yields
\begin{eqnarray}
\sum_{r}\intop_{0}^{\tau}\delta V_{i}\delta V_{j}\left(\frac{\partial^{2}J_{r}}{\partial V_{i}\partial V_{j}}\right)_\tf{eq}\tf{d}t&=&-2\intop_{0}^{\tau}\delta V_{i}\delta V_{j}\left(\frac{\partial I_{i}}{\partial V_{j}}\right)_{\tf{eq}}\tf{d}t, \label{eq:alpha2_case2}
\end{eqnarray}
where we used the Onsager's reciprocity relation $(\partial_{V_j} I_i)_\tf{eq} = (\partial_{V_i} I_j)_\tf{eq}$~\cite{ludovico2016,calvo2017}.
The quantities $(\partial_{V_j} I_i)_\tf{eq}$ are the linear conductances in the limit of small bias voltages.
Therefore this relation shows that these leakage currents, defined as those currents which can not be used to perform any useful work, are also dissipated as heat in the reservoirs, a phenomenon known as Joule heating or Joule law~\cite{ludovico2014,ludovico2016B,ludovico2018B,terrenalonso2019}.
Finally, for $n_\Omega=1$, $n_{V_i}=1$, and $n_{T_k}=0$,
Equation~(\ref{eq:1stLaw_vect}) results in
\begin{eqnarray}
\sum_{r}\intop_{0}^{\tau} \Omega \delta V_{i}\left(\frac{\partial^{2}J_{r}}{\partial{\Omega}\partial V_{i}}\right)_{\tf{eq}}\tf{d}t+\intop_{0}^{\tau}{\Omega}\left(\frac{\partial I_{i}}{\partial{\Omega}}\right)_{\tf{eq}}\delta V_{i}\tf{d}t & = & -\intop_{0}^{\tau}\delta V_{i}\left(\frac{\partial F_{\Omega}}{\partial V_{i}}\right)_{\tf{eq}}{\Omega}\tf{d}t.\label{eq:alpha2_case3}
\end{eqnarray}
Now, using the Onsager's reciprocity relation $(-\partial_{V_i} F_\Omega)_\tf{eq} = (\partial_\Omega I_i)_\tf{eq}$~\cite{fernandez2015,ludovico2016,calvo2017,fernandez2017}, one finds that the first term in the left-hand side of the above equation vanishes, which is an unexpected conservation relation for this second order pumped heat.
We remark that the utility of Equation~(\ref{eq:1stLaw_vect}) relies on the fact that it provides a physical interpretation for the connection between different order contributions that participate in the energy conservation rule.
One can continue analyzing the other relations for $\left|\boldsymbol{\alpha}\right|=2$ and beyond, but the number of relations and cases grow very fast with $\left|\boldsymbol{\alpha}\right|$ and each relation may have its own physical interpretation.
The study of higher-order terms in $|\bm{\alpha}|$ may result useful when addressing particular nonlinear effects in the involved energy currents.
However, for the purpose of the present article we believe that the above analysis is enough as to illustrate the approach proposed by Equation~(\ref{eq:1stLaw_vect}) regarding multi-index expansions.

As the full dynamics of the complete system typically involves a formidable task, many methods in quantum transport treat this problem through a perturbative expansion in $\Omega$.
This is the case of the real-time diagrammatic theory we use in Section~\ref{sec:Coulomb}, where the effective dynamics of the electronic part of the system is described through a perturbative expansion in the characteristic frequency of the driving parameters (which in our case is modeled by the mechanical system), while the voltage and temperature biases are treated exactly.
In this case, the expansion in $\Omega$ comes naturally from the theory itself.
In situations like this, it may result more useful to simplify the expansions by restricting ourselves to those nonequilibrium sources whose perturbative treatment is inherent to the used formalism, as in in~\cite{juergens2013,calvo2017}.

Regarding the above discussion, we now use Equation~(\ref{eq:1stLaw_vect}) to describe how the different energy contributions are linked in an $\Omega$ expansion provided by the theory. The zeroth order terms in $\Omega$ reads
\begin{equation}
\bm{Q}_J^{(0)} \cdot \bm{1} + \bm{Q}_I^{(0)} \cdot \delta \bm{V} = 0.
\end{equation}
This equation shows that the total amount of heat delivered by the leads comes from the bias voltage maintained between them. Its interpretation
is similar to that of Equation~(\ref{eq:alpha2_case2}). The heat term can be associated with the leakage energy current, i.e., the energy flowing
from source to drain leads without being transferred to the local system. The next order in this expansion yields
\begin{equation}
\bm{Q}_J^{(1)} \cdot \bm{1} + \bm{Q}_I^{(1)} \cdot \delta \bm{V} = -\tc{W}_F^{(0)}, \label{eq:general_W=QV}
\end{equation}
and can be understood as a generalization of the motor-pump relation, $\bm{Q}_I^{(1)} \cdot \delta \bm{V} \approx - \tc{W}_F^{(0)}$, discussed in~\cite{bustos2013} for arbitrary bias voltages and temperature gradients. Therefore, according to Equation~(\ref{eq:general_W=QV}), deviations of the mentioned relation at finite voltages are due to the pumped heat induced by the mechanical motion of the local system.
When we evaluate the currents in equilibrium by setting $\delta \bm{V} = 0$ and $\delta \bm{T}=0$, Equation~(\ref{eq:alpha1_case1}) implies 
$\tc{W}_F^{(0)}=0$, meaning that no external work is done in a cycle, and the pumped energy from the leads can again be considered as a leakage current since no net effect on the mechanical system is performed.
If, on the other hand, some bias is present (either thermal or electric), the energy transfer from the leads to the local system imprints a mechanical motion which, in turn, produces some useful work.
The second order term in $\Omega$ gives
\begin{equation}
\bm{Q}_J^{(2)} \cdot \bm{1} + \bm{Q}_I^{(2)} \cdot \delta \bm{V} = -\tc{W}_F^{(1)},
\end{equation}
and generalizes Equation~(\ref{eq:alpha2_case1}) to finite voltage and temperature biases. The right-hand side of Equation~(\ref{eq:alpha2_case1}) can be interpreted as the dissipated energy of the mecanical system, which is delivered to the electronic reservoirs. However, in the above equation 
$\tc{W}_F^{(1)}$ is not guaranteed to be always negative and, from the point of view of the mechanical system, this can be interpreted as a negative friction coefficient.

Now, we return to the multi-variable expansion of the energy currents to establish which orders should be considered in a consistent calculation of the efficiency of quantum motors and pumps. Assuming that in Equation~(\ref{eq:1stLaw_vect}) we take $|\bm{\alpha}|$ up to some truncation value 
$\alpha_\tf{max}$, the order-by-order scheme implies that, for example, the efficiency of the electrically driven quantum motor should be given by
\begin{equation}
\eta = - \frac{\sum_{|\bm{\alpha}| = 0}^{\alpha_\tf{max}} \tc{W}_F^{\bm{\alpha}(\Omega)}}{\sum_{|\bm{\alpha}| = 0}^{\alpha_\tf{max}} 
\bm{Q}_I^{\bm{\alpha}(\bm{V})} \cdot \delta \bm{V}}. \label{eq:eff_gen}
\end{equation}
To illustrate this, let us take the case of a local system coupled to left and right reservoirs at voltages $V_L$ and $V_R$, respectively.
We assume the leads are at the same temperature and we set $\delta V_L = - \delta V_R = \delta V/2$ as the voltage biases. Depending on which kind of expansion we take, one can obtain different expressions for the efficiencies. On the one hand, when expanding in terms of 
$\delta V = V_L-V_R$ and $\Omega$ up to $\alpha_\tf{max} = 2$, the above general expression yields
\begin{equation}
\eta = - \frac{\tc{W}_F^{(0,0)}+\tc{W}_F^{(1,0)}+\tc{W}_F^{(0,1)}}{\left(Q_I^{(0,0)}+Q_I^{(1,0)}+Q_I^{(0,1)}\right) \delta V} = 
-\frac{\tc{W}_F^{(1,0)}+\tc{W}_F^{(0,1)}}{\left(Q_I^{(1,0)}+Q_I^{(0,1)}\right) \delta V},
\end{equation}
where the superscripts indicate the order in the expansions in $\Omega$ and $\delta V$, respectively, and we defined $I = (I_L - I_R)/2$.
As we already mentioned, the zeroth order contributions $\tc{W}_F^{(0,0)}$ and $Q_I^{(0,0)}$ are simply zero as they correspond to the work done by
the conservative part of the CIF and the equilibrium charge current, respectively. On the other hand, when performing an expansion up to $\alpha_\tf{max} = 2$ but only in terms of $\Omega$, Equation~(\ref{eq:eff_gen}) turns into
\begin{equation}
\eta = - \frac{\tc{W}_F^{(0)}+\tc{W}_F^{(1)}}{\left(Q_I^{(0)}+Q_I^{(1)}+Q_I^{(2)}\right) \delta V},
\end{equation}
where now $\tc{W}_F^{(0)}$ and $Q_I^{(0)}$ are nonzero in general, since they are not necessarily evaluated at equilibrium. Although we here restricted ourselves to the efficiency of a nanomotor driven by electric currents, its extension to other operational modes of the device can be obtained from Equation~(\ref{eq:2nd_law}) in a similar way to that of Equation~(\ref{eq:eff_gen}). This procedure then allows us to obtain efficiency expressions for arbitrary expansions in the nonequilibrium sources, which could be useful in the evaluation of the device's performance far from equilibrium.

\section{Quantum motors and pumps in the Coulomb blockade regime} \label{sec:Coulomb}

In this section we consider the CIFs in the so-called Coulomb blockade regime of transport. In this regime, the strong electrostatic repulsion that takes place inside a small quantum dot (usually taken as the local system) highly impacts the device's transport properties, as for small bias voltages no additional charges can flow through the dot and the current gets completely blocked. The full system dynamics in this strongly interacting regime can not be described by, e.g., the scattering matrix approach, and one needs to move to some other theoretical framework. A suitable methodology is given by the real-time diagrammatic theory~\cite{splettstoesser2006}, which allows for an effective treatment of the quantum dot dynamics by performing a double expansion in both the tunnel coupling between the dot and the leads and the frequency associated with the external driving parameters. Since then, many extensions and application examples appeared in the context of quantum pumps~\cite{splettstoesser2008,cavaliere2009,reckermann2010,riwar2010,hiltscher2011,calvo2012,rojek2013,riwar2013,juergens2013,winkler2013,rojek2014,placke2018} and quantum motors~\cite{calvo2017}.

To lowest order in the tunnel coupling, the dot's reduced density matrix obeys the following master equation:
\begin{equation}
\dt{}{t} \bm{p} = \bm{W}\bm{p},
\end{equation}
where the vector $\bm{p} = \{ p_i(t) \}$ describes the dot's occupation probabilities and $\bm{W}$ is the evolution kernel matrix accounting for the 
transition rates between the quantum dot states, due to its coupling to the leads. In the context of CIFs, we assume that the time scale of the mechanical motion, characterized by $\dot{\bm{X}}$, is large as compared to the typical dwell time of the electrons in the local system. This allows for an expansion of the reduced density matrix as $\bm{p} = \sum_{k \geq 0} \bm{p}^{(k)}$, with $\bm{p}^{(k)}$ of order $(\Omega/\Gamma)^k$. Here $\Omega$ and $\Gamma$ denote the characteristic scales for the velocity of the mechanical degrees of freedom and the tunnel rate of the electronic system, respectively, and we always 
assume $\Omega<\Gamma$. The above master equation, in turn, takes the following hierarchical structure~\cite{splettstoesser2006,cavaliere2009}:
\begin{equation}
\bm{W} \bm{p}^{(0)} = \bm{0}, \quad \tf{and} \quad \bm{W} \bm{p}^{(k)} = \frac{\tf{d}}{\tf{d}t} \bm{p}^{(k-1)}.
\end{equation}
In the first equation, $\bm{p}^{(0)}(\bm{X})$ represents the steady-state solution the electronic system arrives when the mechanical system is ``frozen'' at the point $\bm{X}$. As the mechanical coordinate moves in time, this order corresponds to the adiabatic electronic response to the mechanical motion. This has not to be confused with the steady-state regime of the mechanical system, which obviously takes a much longer time to be reached. The second equation contains higher-order nonadiabatic corrections $\bm{p}^{(k)}(\bm{X},\dot{\bm{X}})$ due to retardation effects in the electronic response to the mechanical motion.
In all these equations, the matrix elements of the kernel are of zeroth order in the mechanical velocities, i.e., $\bm{W} = \bm{W}(\bm{X})$. 
Normalization condition on the dot's density matrix implies $\bm{e}^{\tf{T}} \bm{p}^{(k)} = \delta_{k0}$, with $\bm{e}^\tf{T}$ the trace over the dot's Hilbert space. This allows for the definition of an
invertible pseudo-kernel $\tilde{\bm{W}}$, whose matrix elements are $\tilde{W}_{ij} = W_{ij}-W_{ii}$, such that we can write:
\begin{equation}
\bm{p}^{(k)} = \left[ \tilde{\bm{W}}^{-1} \frac{\tf{d}}{\tf{d}t} \right]^k \bm{p}^{(0)}. \label{eq:order-p}
\end{equation}
Once we know the different orders of the reduced density matrix, it is possible to compute the expectation value of any observable $R$ after 
calculation of its corresponding kernel $\bm{W}^R$. This implies the same expansion as before, and it reads
\begin{equation}
R^{(k)} = \bm{e}^{\tf{T}} \bm{W}^{R} \bm{p}^{(k)}. \label{eq:obs_k}
\end{equation}
The observables we are going to address here are the charge and heat currents $I_r$ and $J_r$ associated with the $r$-lead, and defined in Section~\ref{sec:order-by-order}.
In lowest-order in tunneling, and under the assumption that coherences are completely decoupled from the occupations, it is possible to obtain simple expressions for 
the current kernels in terms of the number of particles $n_i$ and energy $E_i$ associated to the dot state $\ket{i}$~\cite{juergens2013,haupt2013}:
\begin{equation}
W^{I_r}_{ij} = -e(n_i-n_j) [W_r]_{ij}, \qquad W^{J_r}_{ij} = -\left[ E_i-E_j-\mu_r (n_i-n_j) \right] [W_r]_{ij}.
\end{equation}
The CIF was derived in Section~\ref{sec:Langevin} under the assumption that the mechanical part of the system, characterized by 
coordinates $\bm{X}$ and associated momenta $\bm{P}$, only interacts with local parameters of the quantum dot through their many-body eigenenergies, 
cf. Equation~(\ref{eq:H_s(e)}). This implies that the $\nu$-component of the CIF operator, defined as $\hat{F}_\nu = -\partial \hat{H}_{\tf{s}}/\partial X_\nu$, is local in the 
quantum dot basis. For this local operator, then, we can simply define a diagonal kernel of the form
\begin{equation}
W^{F_\nu}_{ij} = -\dd{E_i}{X_\nu} \delta_{ij},
\end{equation}
such that Equation~(\ref{eq:obs_k}) gives the $k$-order in the $\Omega$ expansion for any observable.
Importantly, when adding up all contributions from the leads, the above kernel definitions, together with the sum rule $\sum_i W_{ij} = 0$, leads to the following conservation rule for all orders in $\Omega$~\cite{juergens2013}:
\begin{equation}
\sum_r J_r^{(k)} =  - \frac{\tf{d}}{\tf{d}t} \braket{\hat{H}_\tf{s}}^{(k-1)} - \bm{F}^{(k-1)} \cdot \dot{\bm{X}} - \sum_r \frac{\mu_r}{e} I_r^{(k)}.
\end{equation}
This equation, equivalent to~(\ref{eq:E_conservation}), thus expresses the first principle of thermodynamics, which in this case relates the total heat flowing from/into the leads with the variation of the internal energy of the dot and the power contributions due to both mechanical and electrochemical external sources.
By considering a system coupled to two reservoirs $L$ and $R$, with periodic motion characterized by a time period $\tau$, and taking the time integral of the above equation, we recover the frequency expansion of Equation~(\ref{eq:1stLaw_vect}):
\begin{equation}
Q_{J}^{(k)}+Q_{I}^{(k)} \delta V = -\tc{W}_{F}^{(k-1)}, \label{eq:conserv}
\end{equation}
where the bias voltage $\delta V$ is defined through $\mu_L = - \mu_R = e \delta V/2$, $Q_J= Q_{J_L}+Q_{J_R}$, and $Q_I= (Q_{I_L}-Q_{I_R})/2$.
The sign convention employed here implies, for example, that if the left hand side of Equation~(\ref{eq:conserv}) is 
positive, then there is some amount of energy entering into the leads in one cycle, and work is being extracted from the local system.

\subsection{Example: Double quantum dot coupled to a rotor \label{sec:DQD_example}}

In this section we illustrate the discussions of the previous sections in a concrete example based on a double quantum dot (DQD) system locally coupled to a mechanical rotor (see Figure~\ref{fig:1}c). We 
assume a capacitive coupling between the dots and the fixed charges in the rotor such that no charge flows between the two subsystems. The DQD system
is described as in Refs.~\cite{riwar2010,juergens2013,calvo2017}, 
\begin{equation}
\hat{H}_\tf{s} = \sum_{\ell = L, R} \epsilon_\ell \hat{n}_\ell + U \hat{n}_L \hat{n}_R + \frac{U'}{2} \sum_{\ell = L,R} \hat{n}_\ell (\hat{n}_\ell-1)
- \frac{t_c}{2} \left( \hat{d}_{L\sigma}^\dag \hat{d}_{R\sigma} + \tf{h.c.} \right),
\end{equation}
where $\epsilon_\ell$ is the onsite energy and $\hat{n}_\ell = \sum_\sigma \hat{d}_{\ell\sigma}^\dag \hat{d}_{\ell\sigma}$ the number operator 
in the $\ell$-dot, $U$ and $U'$ the inter and intradot charging energies, respectively, and $t_c$ the interdot hopping amplitude. We will work in the 
strong coupling regime $t_c \gg \Gamma$, such that nondiagonal elements (coherences) in the reduced density matrix are decoupled from the diagonal 
ones (occupations), and can be disregarded in first order in tunneling. To simplify the analysis (by reducing the number of states in the two-charge 
block), we work in the limit $U'\rightarrow \infty$, such that double occupation in a single dot is energetically forbidden.

Diagonalization of the above DQD Hamiltonian yields the bonding (b) and antibonding (a) basis for the single-electron charge block, and the reduced density matrix 
reads $\bm{p} = (p_0, p_{\tf{b}\uparrow}, 
p_{\tf{b}\downarrow}, p_{\tf{a}\uparrow}, p_{\tf{a}\downarrow}, p_{\uparrow\uparrow}, p_{\uparrow\downarrow}, p_{\downarrow\uparrow}, 
p_{\downarrow\downarrow})^\tf{T}$. These elements thus denote the probability for the DQD to be either empty ($p_0$), occupied by one electron in the 
$\ell = \tf{b}$ (or $\tf{a}$) orbital with spin $\sigma$ ($p_{\ell \sigma}$), or by two electrons ($p_{\sigma\sigma'}$), one of them in the left dot 
and with spin $\sigma$ and the other electron in the right dot and with spin $\sigma'$. The many-body eigenenergies are therefore $E_0 = 0$ for the 
empty DQD,
\begin{equation}
E_{\tf{b}/\tf{a}} = \frac{\epsilon_L+\epsilon_R}{2} \mp \sqrt{\left(\frac{\epsilon_L-\epsilon_R}{2}\right)^2+\left(\frac{t_c}{2}\right)^2}
\end{equation}
for single-occupation in the bonding or the antibonding orbital, and $E_2 = \epsilon_L+\epsilon_R + U$ for the doubly occupied DQD, respectively.

\begin{figure}[ht]
\begin{centering}
\includegraphics[width=\textwidth]{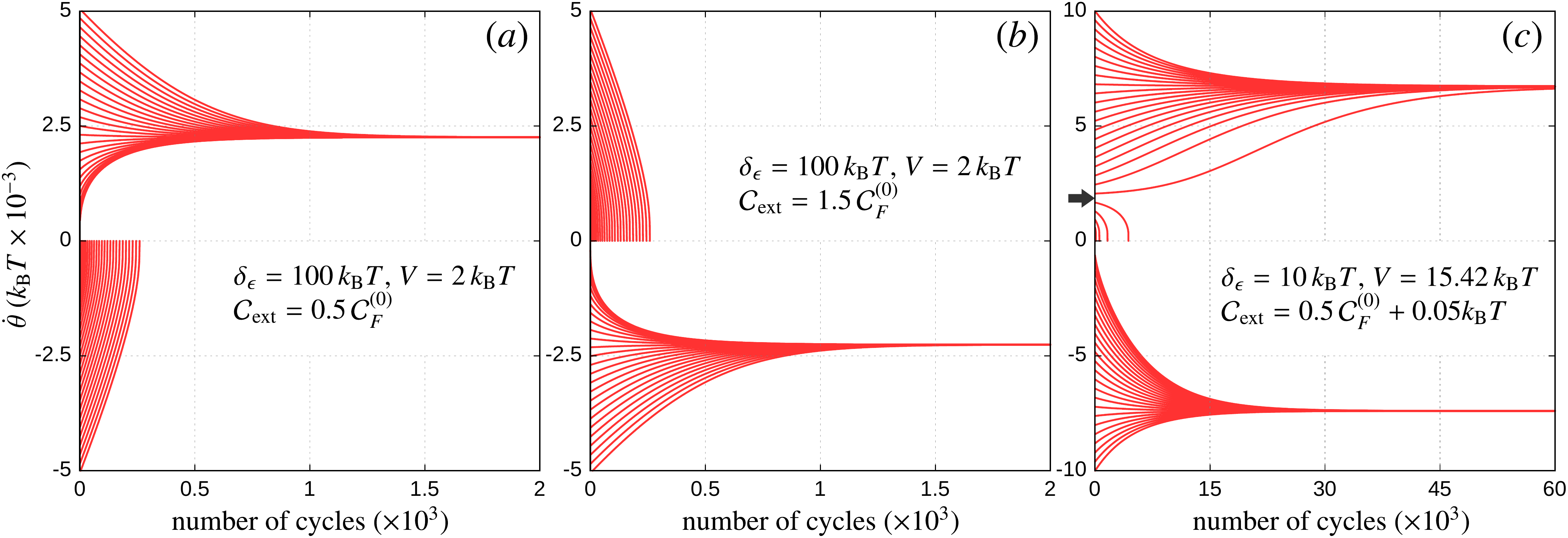}
\end{centering}
\caption{\new{Evolution of the rotor's angular velocity (average over cycle) as a function of the number of cycles for different initial conditions. In panels (\textbf{a}) and (\textbf{b}) we choose $\delta_\epsilon = 100 \, k_\tf{B} T$ and $V = 2 \, k_\tf{B} T$ and we take in (\textbf{a}) $\tc{C}_\tf{ext} = 0.5 \, \tc{C}_F^{(0)}$ so the device operates as a motor, while in (\textbf{b}) we use $\tc{C}_\tf{ext} = 1.5 \, \tc{C}_F^{(0)}$ so the device operates as a pump, cf. Equation~(\ref{eq:vel1}). (\textbf{c}) A situation in which two stable velocities with opposite sign are possible, for $\delta_\epsilon = 10 \, k_\tf{B} T$ and $V = 15.42 \, k_\tf{B} T$ and $\tc{C}_\tf{ext} = \tc{C}_F^{(0)} + 0.05 \, k_\tf{B}T$. The other parameters are: $\delta T = 0$, $U = 20 \, k_\tf{B} T$, $t_c = 10 \, k_\tf{B} T$, $\Gamma_L = \Gamma_R = 0.25 \, k_\tf{B} T$, while for the trajectories we used as working point $\bar{\epsilon}_L = \bar{\epsilon}_R = -6 \, k_\tf{B} T + \delta_\epsilon/\sqrt{2}$ .}} 
\label{fig:2}
\end{figure}

To account for the coupling between electronic and mechanical degrees of freedom we take as mechanical coordinate the angle $\theta$ describing the orientation of the rotor axis. We assume the following dependence through the onsite energies:
\begin{equation}
\epsilon_L(\theta) = \bar{\epsilon}_L + \lambda x_L = \bar{\epsilon}_L + \delta_\epsilon \cos(\theta), \quad \tf{and} \quad
\epsilon_R(\theta) = \bar{\epsilon}_R + \lambda x_R = \bar{\epsilon}_R + \delta_\epsilon \sin(\theta),
\end{equation}
which defines a circular trajectory in energy space of radius $\delta_\epsilon = \lambda r_0$ ($r_0$ measures the actual radius of the rotor) around 
the working point $(\bar{\epsilon}_L,\bar{\epsilon}_R)$. For simplicity in what follows, we will focus in the tangential component of the force only, 
by assuming that the radial component is always compensated by internal forces in the rotor (e.g., fixed charges along the rotor's axis). By 
applying this form in Equation~(\ref{eq:lang}), we obtain the equation of motion~(\ref{eq:lang_ang}) for the angular velocity $\dot{\theta}$, where: $\tc{I}=m_\tf{eff} r_0^2$ is the rotor's moment of inertia, $\tc{F} = \sum_i (-\partial_\theta E_i) p_i = r_0 \bm{F}\cdot 
\hat{\boldsymbol{\theta}}$ is the current-induced torque, $\tc{F}_\tf{ext}$ is the \new{torque produced by the external force}, which is assumed to be constant along the whole trajectory, and $\xi_\theta$ accounts for the force fluctuations. The stationary regime is reached once the rotor performs a periodic motion (characterized by a time period $\tau$) with no overall acceleration, i.e., when Equation~(\ref{eq:s-s_condition}) is fulfilled. \new{We show in Figure~\ref{fig:2} some examples of the evolution of the rotor's angular velocity for different initial conditions. This is plotted as a function of the number of cycles performed by the rotor, instead of time, to unify scaling along the horizontal axis. Here we take a sufficiently large moment of inertia such that the variation $\Delta\dot\theta$ over one period is small as compared to the averaged value of $\dot\theta$ in one cycle. This, in turn, implies a large number of cycles for the rotor to reach the steady-state regime.}

As discussed in Section~\ref{sec:steady_state}, when we take $\tc{I}$ sufficiently large $\Delta\dot{\theta} \rightarrow 0$, and it is possible to approximate $\dot\theta$ as constant during the whole cycle. This allows us to take out this quantity from the integrals defining the different orders of the energy currents, as we did in Equation~(\ref{eq:const_vel_condit}). The problem is, however, that the sign in $\dot{\theta}$ is not 
\textit{a priori} known and, with it, the direction of the 
trajectory for the line integrals defining $\tc{W}_{F}^{(k)}$. To become independent of this issue, we denote by $s$ the sign of $\dot{\theta}$, and notice that $\tc{F}^{(k)} = f_\theta^{(k)} \dot{\theta}^k$, where
$f_\theta^{(k)} = \partial_{\dot{\theta}}^k \tc{F} |_{\dot{\theta}=0}/k!$
only depends parametrically on $\theta$, so we have
\begin{equation}
\tc{W}_{F}^{(k)} = \oint_{C(s)} \bm{F}^{(k)}\cdot \tf{d}\bm{X}
= \int_0^{2\pi s} \tc{F}^{(k)} \tf{d}\theta
= s \left[ \int_0^{2\pi} f_\theta^{(k)} \tf{d}\theta \right] \dot{\theta}^k
= s \tc{C}_F^{(k)} \dot{\theta}^k,
\label{eq:coeff}
\end{equation}
where, importantly, the coefficients $\tc{C}_F^{(k)}$ are independent on $s$, so the $\dot{\theta}^k$ and the extra sign $s$ accompanying 
$\tc{C}_F^{(k)}$ in the right hand side of the equation give the correct sign for $\tc{W}_{F}^{(k)}$. As discussed in Section~\ref{sec:steady_state}, the steady-state condition in Equation~(\ref{eq:s-s_condition}) can be viewed as the equation for the final velocity that the rotor acquires as a function of the \new{external} force, which in the present case turns into:
\begin{equation}
0 = \sum_k \tc{C}_F^{(k)} \dot{\theta}^{k} - \tc{C}_\tf{ext}, \label{eq:velocity}
\end{equation}
where we defined $\tc{C}_\tf{ext} = 2\pi \tc{F}_\tf{ext}$.
The stability of the final solution is inherited from Equation~(\ref{eq:lang_ang}) 
and, in our case with $\dot{\theta}$ constant, it is given by
\begin{equation}
\sum_k \tc{C}_F^{(k)} k \dot{\theta}^{k-1} < 0. \label{eq:stab}
\end{equation}
Notice the specific case where $\tc{C}_\tf{ext} = \tc{C}_F^{(0)}$, i.e., the \new{external} work equals the bias work coming 
from the first order currents, cf. Equation~(\ref{eq:conserv}) for $k=1$. In this case there is always a trivial solution, given by 
$\dot{\theta} = 0$. Of course, this solution is useless as the system becomes frozen, and there is no energy transfer between the leads 
and the mechanical system. However, this situation marks one of the transition points between the operation modes of the device. If we now 
consider an infinitesimal difference between these two coefficients, we can faithfully treat Equation~(\ref{eq:velocity}) up to linear orden in 
$\dot{\theta}$, such that the solution is
\begin{equation}
\dot{\theta} = \frac{\tc{C}_\tf{ext}-\tc{C}_F^{(0)}}{\tc{C}_F^{(1)}}, \label{eq:vel1}
\end{equation}
together with the condition $\tc{C}_F^{(1)}<0$, in agreement with Equation~(\ref{eq:stab}). This implies that the integral of the friction coefficient 
over a cycle needs to be positive, otherwise the rotor can not reach the stationary regime. As the sign in 
$\tc{C}_F^{(1)}$ is fixed to this order, the sign in $\dot{\theta}$ only depends on the relation between $\tc{C}_\tf{ext}$ and 
$\tc{C}_F^{(0)}$. This, together with the stationary condition $\tc{W}_\tf{ext} = \tc{W}_F$, establishes the operation mode of the 
device, in the sense that with this information we can deduce the sign of $\tc{W}_\tf{ext} =  s \tc{C}_\tf{ext}$, and hence the direction 
of the energy flow. If, for example, $\tc{W}_\tf{ext} < 0$, the external energy is delivered into the DQD, which in turn it goes as energy 
current to the leads, so the device acts as an energy pump. On the other hand, if $\tc{W}_\tf{ext} > 0$ the energy current coming from the leads goes into the DQD and it is transformed into mechanical work, so the device operates as a motor. \new{These two scenarios are shown in Figure~\ref{fig:2}, where we take $\tc{C}_\tf{ext}>0$, such that the positive sign in the final velocity implies that the device acts as a motor (see panel a), while a negative sign implies that the device operates as a pump (see panel b).}
Of course, when regarding the device as an energy pump, the above criterion only establishes those regions in $\tc{C}_\tf{ext}$ where we can expect some kind of pumping mechanism. Depending 
then on the specific type of pumping we have in mind, the ranges in $\tc{C}_\tf{ext}$ will be subject to additional conditions.
For example, if we would like to have this device acting as a quantum charge pump, then we should check those regimes in $\tc{C}_\tf{ext}$ where the electrons flow against the bias voltage. This discussion will be reserved to the next section, and for now we will only define the operation mode from the direction of the energy flow.

Let us now consider second-order effects due to $\tc{C}_F^{(2)}$ in Equation~(\ref{eq:velocity}), where we obtain:
\begin{equation}
\dot{\theta}_\pm = -\frac{\tc{C}_F^{(1)}}{2\tc{C}_F^{(2)}} \pm 
\sqrt{\left( \frac{\tc{C}_F^{(1)}}{2\tc{C}_F^{(2)}} \right)^2+\frac{\tc{C}_\tf{ext}-\tc{C}_F^{(0)}}{\tc{C}_F^{(2)}}},
\label{eq:vel2}
\end{equation}
together with the conditions:
\begin{equation}
\left( \frac{\tc{C}_F^{(1)}}{2\tc{C}_F^{(2)}} \right)^2 > \frac{\tc{C}_F^{(0)}-\tc{C}_\tf{ext}}{\tc{C}_F^{(2)}}, \qquad 
\pm \tc{C}_F^{(2)} < 0.
\end{equation}
The first inequality ensures a positive argument in the square root of Equation~(\ref{eq:vel2}), so $\dot{\theta}$ is a real number, while the second 
inequality comes from the stability condition given by Equation~(\ref{eq:stab}), and tells us which branch one should choose in Equation~(\ref{eq:vel2}) to get the stable solution: If $\tc{C}_F^{(2)}>0$ then $\dot{\theta}_{-}$, and if $\tc{C}_F^{(2)}<0$ then $\dot{\theta}_{+}$. Importantly, far from 
equilibrium (e.g., $\delta V \gg k_\tf{B} T$) it is possible to arrive to the odd situation where $\tc{C}_F^{(1)}>0$, and the ``dissipated'' energy 
$\tc{W}_F^{(1)} = s \tc{C}_F^{(1)} \dot{\theta} = \tc{C}_F^{(1)} |\dot{\theta}|$ is positive. This, however, does not prevent the 
rotor to reach the stationary regime (to this order in $\dot{\theta}$), since the lower-order terms compensate this energy gain. Therefore, one can end up in a situation where the second-order energy current comes from the leads and it is delivered into the mechanical system (thus favouring the motor regime), contrary to the standard situation where the ``dissipated'' energy flows to the leads. Regarding Equation~(\ref{eq:vel2}), the sign of $\dot{\theta}$ now depends on the relation between the first term and the square root, together with the sign of $\tc{C}_F^{(2)}$, so this analysis is not that simple as in the linear case. However, once we have the $\tc{C}_F$-coefficients we can always determine if the rotor is able to reach the stationary regime and infer whether $\tc{W}_\tf{ext}$ is positive or negative and, with it, the operation mode of the device.

For higher orders in $\dot{\theta}$, the above analysis for the operation mode of the device is the same, but more ingredients 
may come into play due to the (order-dependent) specific solutions for the stationary value of $\dot{\theta}$. Interestingly, by including 
higher-order terms in Equation~(\ref{eq:velocity}), it could happen that for a fixed choice of the parameters ($\tc{F}_\tf{ext}$, $\delta V$, $\delta T$, etc.) the system presents more than one stable solution, and even with different signs in $\tc{W}_\tf{ext}$, so the initial condition on 
$\dot{\theta}$ decides the operation mode of the device once the rotor reaches the stationary regime. \new{In Figure~\ref{fig:2}c we show this nolinear effect in the dynamics of the rotor's velocity. This was done by taking $\tc{F}$ up to third order in the $\dot{\theta}$ expansion, where we find two stable solutions ($\dot{\theta} \sim -7.4 \times 10^{-3} \, k_\tf{B}T$ and $6.7 \times 10^{-3} \, k_\tf{B}T$ ) and one unstable solution in between ($\dot{\theta} = 2 \times 10^{-3} \, k_\tf{B}T$). Notice that, in this case, if the rotor starts with a positive initial condition below the unstable solution (see gray arrow), then it can not reach the negative stable solution, as once $\dot\theta = 0$ the rotor gets trapped in a local minimum.}

\begin{figure}[ht]
\begin{center}
\includegraphics[width=.8\textwidth]{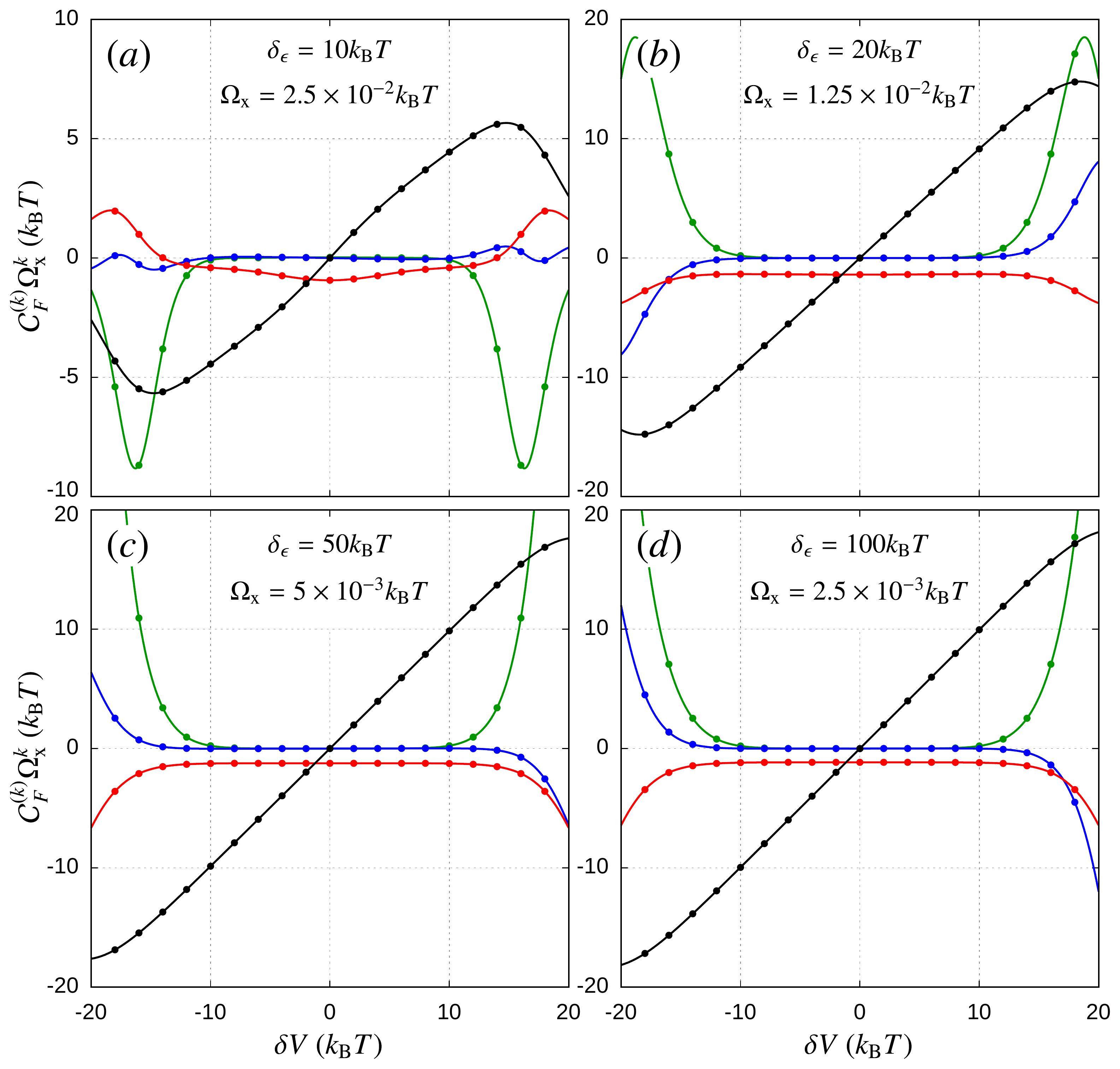}
\end{center}
\caption{\new{Order-by-order contributions to the current-induced work as a function of the bias voltage for different orbit radius: 
(\textbf{a}) $\delta_\epsilon = 10$ and $\Omega_\tf{x} = 2.5 \times 10^{-2}$, (\textbf{b}) $\delta_\epsilon = 20$ and $\Omega_\tf{x} = 1.25 \times 10^{-2}$, (\textbf{c}) $\delta_\epsilon = 50$ and $\Omega_\tf{x} = 5 \times 10^{-3}$, (\textbf{d}) $\delta_\epsilon = 100$ and $\Omega_\tf{x} = 2.5 \times 10^{-3}$, in units of $k_\tf{B} T$. We show, in solid lines, the zeroth (black), first (red), second (blue) and third (green) orders in $\dot{\theta}$, respectively. The circles accompanying the lines correspond to $(\tc{C}_J^{(k)}+\tc{C}_I^{(k)}\delta V)\Omega_x^{(k-1)}$, 
with $\tc{C}_I^{(k)}$ and $\tc{C}_J^{(k)}$ defined in Equation~(\ref{eq:current-coeff}), and numerically confirm Equations~(\ref{eq:conserv}) and (\ref{eq:cons-coeff}). The remaining parameters coincide with those of Figure~\ref{fig:2}.}}
\label{fig:3}
\end{figure}

\new{In Figure~\ref{fig:3} we show how the $\tc{C}_F$-coefficients contribute to the current-induced work as a function of the bias voltage for different values of the orbit radius $\delta_\epsilon$. According to Equation~(\ref{eq:coeff}), these coefficients can only be compared upon multiplication with the $k$-power of some frequency of reference $\Omega_\tf{x}$. We assess this frequency of reference by performing the following analysis: From Equation~(\ref{eq:order-p}) we can estimate the maximum allowed value for $\dot{\theta}$ compatible with the frequency expansion. Since $\tilde{\bm{W}}^{-1} \propto \Gamma^{-1}$, with $\Gamma = \Gamma_L + \Gamma_R$, and
\begin{equation}
\dt{}{t} = \dot{\theta} \dd{}{\theta} = \dot{\theta} \sum_i \dd{\epsilon_i}{\theta} \dd{}{\epsilon_i},
\end{equation}
we obtain
\begin{equation}
\bm{p}^{(k)} = \left[ \tilde{\bm{W}}^{-1} \dot{\theta} \sum_i \dd{\epsilon_i}{\theta} \dd{}{\epsilon_i} \right]^k \bm{p}^{(0)}
\propto  \left[ \frac{\Omega}{\Gamma} \frac{\delta_\epsilon}{k_\tf{B}T} \right]^k,
\end{equation}
where we use the fact that $\partial_\theta \epsilon_i \propto \delta_\epsilon$ and the energy derivative of both $\bm{p}^{(0)}$ and $\tilde{\bm{W}}^{-1}$ are proportional to $1/k_\tf{B}T$. As the consistence of the frequency expansion relies on the convergence of the occupations, this yields the adiabaticity condition~\cite{juergens2013,calvo2017}
\begin{equation}
\frac{\Omega}{\Gamma} \ll \frac{k_\tf{B}T}{\delta_\epsilon},
\end{equation}
from which we can estimate the maximum allowed frecuency $\Omega_\tf{max}$ as
\begin{equation}
\Omega_\tf{max} = \frac{k_\tf{B}T}{\delta_\epsilon} \Gamma.
\end{equation}
Obviously, this extreme value sets the point where the expansion could diverge, so to illustrate the different order contributions to the current-induced work we take, in Figure~\ref{fig:3}, an intermediate referece frequency $\Omega_\tf{x} = \Omega_\tf{max}/2$.
As we can see, for a long range of the bias voltage ($V \lesssim 10 \, k_\tf{B}T$), both the zeroth and first order terms contribute, while higher-order terms are almost negligible. The zeroth order contributions (black) show a linear dependence whose slope increases with the orbit size, up to some saturation value, related with the quantization of the pumped charge. The first order terms (red) remain almost constant and negative in this bias regime. This implies that the linear order treatment given in Equation~(\ref{eq:vel1}) is enough, as long as $\tc{C}_\tf{ext} \sim \tc{C}_F^{(0)}$. For $V \gtrsim 10 \, k_\tf{B}T$, we can observe non-equilibrium effects like ``inverse dissipation'', i.e. a positive contribution from the first order term (red) in Figure~\ref{fig:3}a. Additionally, the higher-order terms (blue and green) can become larger than the first two and they need to be included in the calculation of the final velocity through Equation~(\ref{eq:velocity}) or in the current-induced force appearing in Equation~(\ref{eq:lang_ang}). This, in turn, could lead to several stable solutions whose validity should be determined through a systematic convergence analysis that is beyond the scope of the present work. For the particular choice of parameters used in Figure~\ref{fig:2}c we checked that the next-order coefficient ($\tc{C}_F^{(4)}$) has a negligible impact on the third order solutions.}

\new{To clarify the role of $\tc{C}_F$-coefficients in the operation regime of the device, let us take for example $\delta_\epsilon = 100 \, k_\tf{B} T$ (see Figure~\ref{fig:3}d) and $\delta V \sim 2 \, k_\tf{B}T$. As the sign in $\tc{C}_F^{(1)}$ is negative, the linear order solution given by Equation~(\ref{eq:vel1}) is stable. If we start with 
$\tc{C}_\tf{ext} = 0$, then $\dot{\theta}$ is positive, so for $0 < \tc{C}_\tf{ext} < \tc{C}_F^{(0)}$ the device acts as a motor (see Figure~\ref{fig:2}a). When $\tc{C}_\tf{ext} > \tc{C}_F^{(0)} > 0$, $\dot{\theta}$ becomes negative and hence the device operates as a pump (see Figure~\ref{fig:2}b). Additionally, when $\tc{C}_\tf{ext} < 0$ the angular velocity is still positive, but as we changed the sign in $\tc{C}_\tf{ext}$ we have that $\tc{W}_\tf{ext} < 0$, so the device operates as a pump. This is the other transition point between the two operation modes of the device, i.e., the sign in $\tc{W}_\tf{ext} = s \tc{C}_\tf{ext}$ changes with $s$ and the sign of $\tc{C}_\tf{ext}$. It is also interesting to notice how the operation regimes change with the sign of the bias voltage. For the force coefficients, we can see in the figure that they have definite parity with respect to the bias voltage
\begin{equation}
\tc{C}_F^{(k)}(-\delta V) = (-1)^{k+1} \tc{C}_F^{(k)}(\delta V), \label{eq:cF-v1}
\end{equation}
since $\Omega_\tf{x}$ remains constant for a fixed orbit radius. For the chosen parameters $\bar{\epsilon}_L = \bar{\epsilon}_R$, $\Gamma_L = \Gamma_R$, and $\delta T = 0$, the transformation $\delta V \rightarrow - \delta V$ can be regarded as the inversion operation $(L,R) \rightarrow (R,L)$, which changes the sign of $\dot{\theta}$, i.e., $\dot{\theta}(-\delta V) = - \dot{\theta}(\delta V)$. In this sense, if there is some finite temperature gradient, this operation should also involve the sign inversion of $\delta T$. \new{To infer how this bias inversion affects the final velocity of the device, we can replace Equation~(\ref{eq:cF-v1}) in Equation~(\ref{eq:velocity}). We can therefore recognize the transformation $\dot\theta (-\delta V) = -\dot\theta(\delta V)$ if we also change the sign of the external force, such that $\tc{C}_\tf{ext}(-\delta V) = - \tc{C}_\tf{ext}(\delta V)$.} The even/odd parity in the $k$-order coefficient thus implies that the current-induced work is invariant under such a transformation, cf. Equation~(\ref{eq:coeff}), and the ranges for the operation regimes of the device remain the same if we invert the sign of the external force. Of course, this analysis is no longer valid in more general situations where $\bar{\epsilon}_L \neq \bar{\epsilon}_R$ or $\Gamma_L \neq \Gamma_R$, such that the change in the sign of $\delta V$ (or $\delta T$) can not be related with the left-right inversion operation.}

As an additional test for Equations~(\ref{eq:conserv}) and (\ref{eq:1stLaw_vect}), we define equivalent coefficients for the amount of transported charge and heat in a cycle
\begin{equation}
\tc{C}_I^{(k)} = \int_0^{2\pi} \left.\dd{^k I_\theta}{\dot{\theta}^k}\right|_{\dot{\theta}=0} \frac{\tf{d}\theta}{k!}, \qquad
\tc{C}_J^{(k)} = \int_0^{2\pi} \left.\dd{^k J_\theta}{\dot{\theta}^k}\right|_{\dot{\theta}=0} \frac{\tf{d}\theta}{k!},
\label{eq:current-coeff}
\end{equation}
such that $Q_I^{(k)} = s \tc{C}_I^{(k)} \dot{\theta}^{k-1}$, and $Q_J^{(k)} = s \tc{C}_J^{(k)} \dot{\theta}^{k-1}$. These contributions can be 
evaluated independently from the $\tc{C}_F$-coefficients and in Figure~\ref{fig:3} these are shown in circles, which gives numerical agreement for the 
energy conservation principle
\begin{equation}
\tc{C}_J^{(k)}+\tc{C}_I^{(k)} \delta V = -\tc{C}_F^{(k-1)}, \label{eq:cons-coeff}
\end{equation}
in the considered orders of $\dot{\theta}$.
For the considered example in Figure~\ref{fig:3}, the current coefficients $\tc{C}_I$ and $\tc{C}_J$ also show 
(independently) a definite parity with respect to the bias voltage. In fact, as Equation~(\ref{eq:cons-coeff}) suggests, the heat current coefficients 
$\tc{C}_J^{(k)}$ present the same parity as $\tc{C}_F^{(k-1)}$, while the charge current coefficients $\tc{C}_I^{(k)}$ need to have the opposite 
parity since they are multiplied by $\delta V$.

\subsubsection{Motor-pump efficiencies}

\begin{figure}
\begin{centering}
\includegraphics[width=0.9\textwidth]{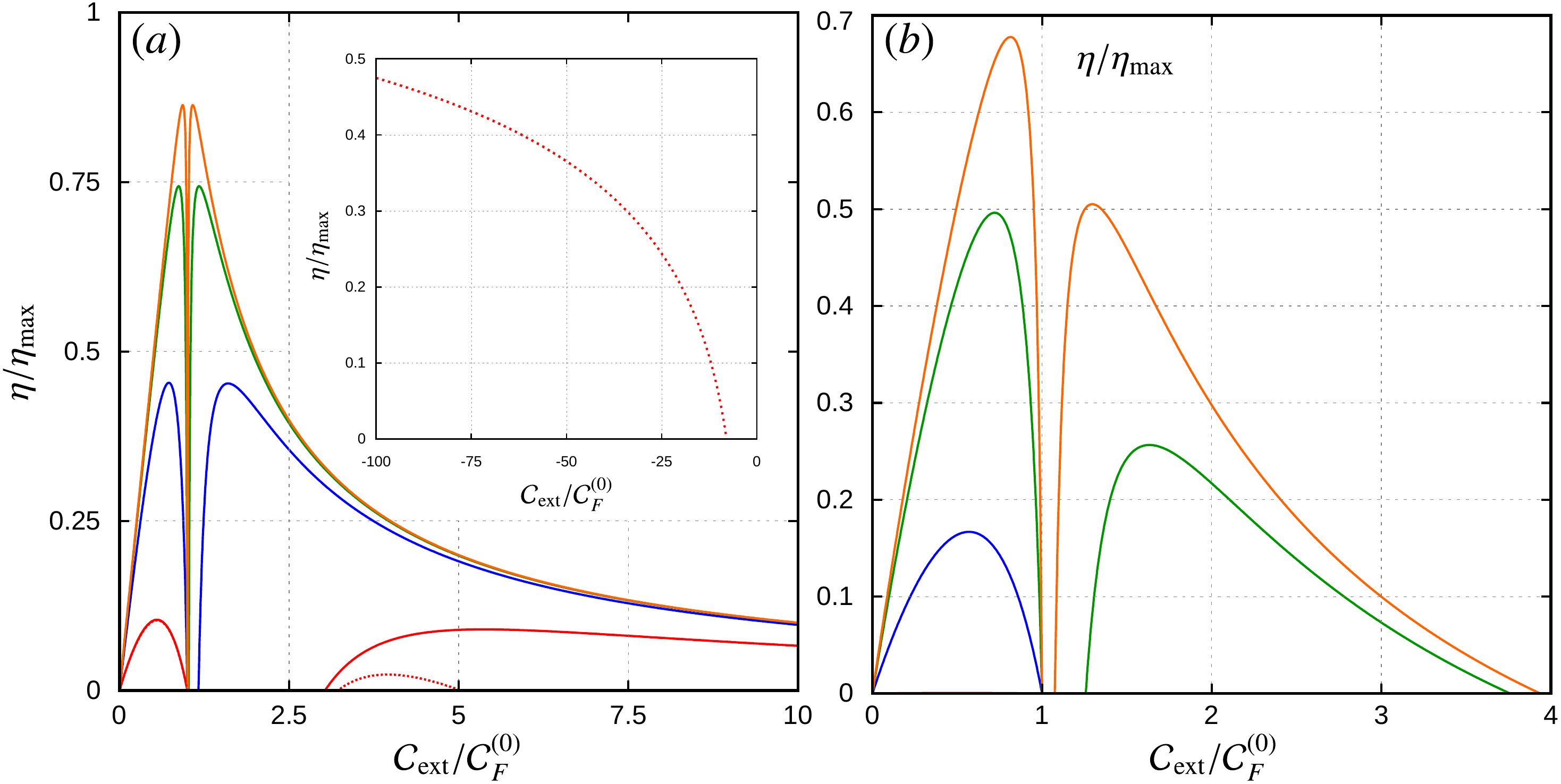}
\end{centering}
\caption{
Normalized efficiencies as a function of the \new{normalized external work ($\tc{C}_\tf{ext}/\tc{C}_F^{(0)}$)} and orbit radius: $\delta_\epsilon = 10$ (red), $20$ (blue), $50$ (green), 
and $100$ (orange), in units of $k_\tf{B} T$.
Panel (\textbf{a}) shows the electric motor/pump lowest order efficiencies (solid lines) for the device driven by a bias voltage $\delta V = 2 k_\tf{B} T$ and $\delta T = 0$. The next-order efficiency for the smallest orbit is shown in dotted red and in the inset for negative $\tc{C}_\tf{ext}$. Panel (\textbf{b}) shows the heat engine/refrigerator lowest order efficiencies for the device driven by a temperature gradient $\delta T = 0.5 T$ and $\delta V = 0$. The other parameters coincide with those of Figure~\ref{fig:2}.
}
\label{fig:4}
\end{figure}

As we stated above, the sign of the \new{external} force, together with the rotor's stationary condition and the first law of thermodynamics, 
determine the direction of the energy flow and, with it, the operation mode of the device. Obviously, as no other power sources are involved in 
this example, the efficiency of this energy conversion, defined as $\eta =$ (output power)/(input power), is always equal to 1.
This, however, only establishes those regions in the parameter space where we can expect the device operating as a quantum motor or a quantum energy pump. In this section we discuss the particular conditions that appear when the device operates through a specific type of current. In this sense, the motor regime corresponds to the situation in which a transport current (e.g., charge, heat, spin, etc.) flowing through the leads in response to a bias (voltage, temperature, spin polarization, etc.) delivers some amount of energy into the local system which can be used as mechanical work. The pump regime, on the other hand, corresponds to the inverse operation in which the \new{external} work is exploited to produce a current flowing against the imposed bias. This topic was also discussed in~\cite{juergens2013} for charge and heat currents in a DQD device, where limitations to the efficiency of the considered processes were attributed to the different orders appearing in the frequency expansion of the currents. We here provide a similar analysis in terms of our explicit model for the mechanical system. The \new{inclusion of the external force in the description of the model}, as we shall see next, appears as the key ingredient in bridging the motor and pump regimes for a given choice of the bias.

For the device acting as a motor, the output power should be given by $\tc{W}_\tf{ext}/\tau$, under the condition $\tc{W}_\tf{ext} > 0$, but we still need to specify the input power. If we consider that the mechanical rotor is driven by the electric current, i.e., due to some applied bias voltage and no thermal gradient applied, the input power is given by $-\bm{Q}_I\cdot \delta \bm{V}/\tau$, and hence the efficiency of this type of motor is
\begin{equation}
\eta = -\frac{\tc{W}_\tf{ext}}{\bm{Q}_I \cdot \delta \bm{V}} = 1 + \frac{Q_J}{Q_I \delta V}.
\end{equation}
Equation~(\ref{eq:electric-motor}) establishes that the maximum efficiency for this device is $\eta \leq 1$, and the above equation tells us that 
the heat current produced by the bias voltage reduces the motor's performance. As in this work we calculate such currents through an expansion
in the angular velocity, the efficiency is also limited by this expansion.
If in the calculation of $\dot{\theta}$ we consider Equation~(\ref{eq:velocity}) up to first order in $\tc{C}_F^{(k)}$, then, as discussed in Section~\ref{sec:p-m_rel}, order-by-order energy conservation demands that the currents are to be considered up to second order, and in terms of the current coefficients this takes the form:
\begin{equation}
\eta = -\frac{\tc{C}_\tf{ext}/\delta V}{\tc{C}_I^{(0)} \dot{\theta}^{-1}+ \tc{C}_I^{(1)} + \tc{C}_I^{(2)} \dot{\theta}}. \label{eq:hc-m}
\end{equation}
In the limits of the motor's operation regime, given by $\tc{C}_\tf{ext} = 0$ and $\tc{C}_\tf{ext} = \tc{C}_F^{(0)}$, it is easy to 
see that the efficiency goes to zero, since for $\tc{C}_\tf{ext} = 0$ the numerator in the above expression is zero, and for 
$\tc{C}_\tf{ext} = \tc{C}_F^{(0)}$ the rotor's velocity goes to zero, so the denominator grows to infinity due to the contribution 
$\tc{C}_I^{(0)}/\dot{\theta}$ from the leakage current. The same happens if we consistently include higher-order terms in this expression. For example,
if we use Equation~(\ref{eq:vel2}) for the rotor's angular velocity, then we should add $\tc{C}_I^{(3)} \dot{\theta}^2$ in the denominator.

\begin{figure}[ht]
\begin{centering}
\includegraphics[width=\textwidth]{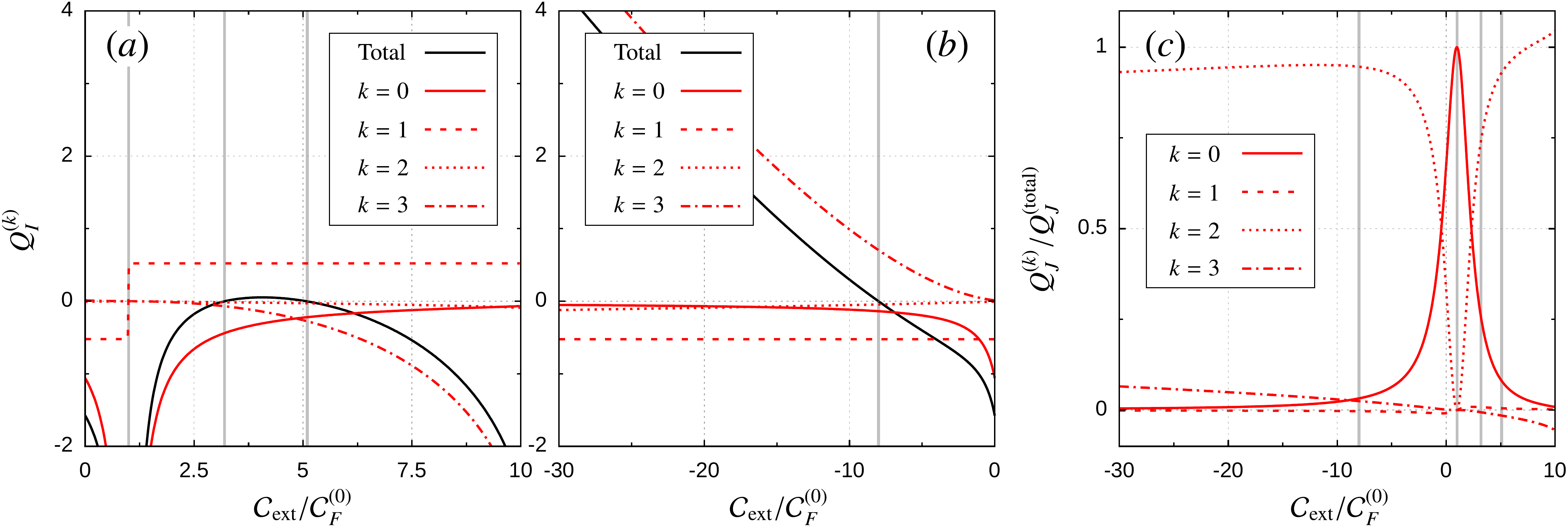}
\end{centering}
\caption{
\new{
Different order contributions to the pumped charge and heat as a function of the normalized external work for the case $\delta_\epsilon = 10 \, k_\tf{B}T$ (dotted red in Figure~\ref{fig:4}a).
Panels (\textbf{a}) and (\textbf{b}) show the $k$-th order pumped charge $Q_I^{(k)} = (Q_{I_L}^{(k)} - Q_{I_R}^{(k)})/2$. The sum of all these contributions is shown in solid black and denoted as $Q_I^{(\tf{total})}$. (\textbf{c}) Pumped heat contributions to left and right reservoirs $Q_{J}^{(k)} = Q_{J_L}^{(k)} + Q_{J_R}^{(k)}$, divided by $Q_J^{(\tf{total})}$, i.e. the sum of all contributions from $k = 0$ to $k = 3$. The vertical gray lines mark different transition points: $\tc{C}_\tf{ext}/\tc{C}_F^{(0)}=1$ is the motor/pump (energy pump) transition, while the other lines correspond to transitions between frustrated-pump/charge-pump regimes, i.e., when $Q_I^\tf{(total)}$ changes its sign.
}
}
\label{fig:5}
\end{figure}

Away from this region, we enter in the ``pumping domain'' characterized by a charge current which opposes to the ``natural'' direction dictated by 
the bias voltage, and $\tc{W}_\tf{ext}<0$. In this sense, the input power and the output power are inverted with respect to the motor region and,
consequently, the efficiency of this ``battery charger'' device is given by
\begin{equation}
\eta = -\frac{\tc{C}_I^{(0)} \dot{\theta}^{-1}+ \tc{C}_I^{(1)} + \tc{C}_I^{(2)} \dot{\theta}}{\tc{C}_\tf{ext}/ \delta V}. \label{eq:hc-p}
\end{equation}
In this regime there is, however, an additional condition to be fulfilled, which is $Q_I \delta V > 0$. Regarding the different orders in $Q_I$, it usually  happens that close to the transition point $\tc{C}_\tf{ext} = \tc{C}_F^{(0)}$ the charge current still flows in the bias direction, since it is dominated by the leakage current. In this region we say that the pumping mechanism is ``frustrated'' as the energy delivered by the rotor is not 
enough as to reverse the direction of the charge current. Going away from this region the angular velocity acquires some finite value, reducing the 
zeroth-order contribution $\tc{C}_I^{(0)}/\dot{\theta}$ to the point where it is equal to the higher-order contributions 
$\tc{C}_I^{(1)} + \tc{C}_I^{(2)}\dot{\theta}$, thus marking the activation point of the charge pump. In Figure~\ref{fig:4}a we show the efficiency of the device as a function of the \new{external} force for different orbit sizes and fixed bias $\delta V = 2 k_\tf{B}T$. In all cases, the device starts from $\tc{C}_\tf{ext} = 0$ as a motor and its efficiency reaches a maximum which increases with the orbit size. Soon after this point the motor's efficiency decreases to zero due to the leakage current effect, which becomes dominant at $\tc{C}_\tf{ext} = \tc{C}_F^{(0)}$.
From this point we can observe the gapped region for the frustrated pump, which is more pronounced for small orbits, since the first-order pumped charge $Q_I^{(1)}$ is smaller than its quantized limit and thereby it takes a larger value of $\tc{C}_\tf{ext}$ to compensate the amount of pumped leakage current $Q_I^{(0)}$ in a cycle.

\new{Up to this point, we have discussed only the effect of the lowest-order terms of the expansion in $\dot\theta$, given by Equations~(\ref{eq:vel1}),~(\ref{eq:hc-m}) and (\ref{eq:hc-p}). For the smallest orbit, in addition, we show in dotted red line the next-order efficiency, obtained from Equation~(\ref{eq:vel2}) and adding the $Q_I^{(3)}$ term to Equations ~(\ref{eq:hc-m}) and~(\ref{eq:hc-p}). We can see that for the motor regime there are no significant changes, but for the pump regime important differences appear.
Firstly, in the $\tc{C}_\tf{ext}>0$ region, there is a cut-off for the external force in which the pumping mechanism is again frustrated, i.e., the charge current again points in the bias direction. As can be seen in Figures~\ref{fig:5}a and c, this decreasing in the efficiency is not  attributed to the extra heat dissipated to the reservoirs, as one may expect. Note in the figure that the extra contribution to the pumped heat, $Q_J^{(3)}$, is negligible as compared to the lowest order terms. What happens here is that the third-order contribution to the pumped charge $Q_I^{(3)}$ rapidly becomes dominant in the charge pump region, causing a sudden drop in the efficiency and, with it, the appearance of a second frustrated-pump region. Secondly, another higher-order effect appears in the $\tc{C}_\tf{ext}<0$ region. There, the efficiency is nonzero for $\tc{C}_\tf{ext}/\tc{C}_F^{(0)} < -8$ (see inset in Figure~\ref{fig:4}a), meaning that the pumping mechanism can be activated even when the external force points in the same direction as that of the current-induced force.~\footnote{Given the convention used for $\tc{F}_\tf{ext}$ in Equation~(\ref{eq:lang}) and the chosen parameters, in this region the sign of $\dot\theta$ remains the same as that when $\tc{F}_\tf{ext} = 0$. There, the zeroth and first order contributions to the charge current flow in the same direction.} 
Again, the third-order term is the one that reverses the direction of the total charge current, see Figure~\ref{fig:5}b. It is important to mention that the purpose of the present discussion is only to highlight deviations from the linear solution of Equation~(\ref{eq:vel1}), not to analyse the convergence of the total pumped current. For the larger values of $\delta_\epsilon$ used in Figure~\ref{fig:4}a we do not show the next order corrections, as they are negligible in the shown range of $\tc{C}_\tf{ext}$.
}

An analysis similar to the above one can be carried out for the device driven by a temperature gradient $\delta T$, defined through $T_L = T+\delta T/2$ and $T_R = T-\delta T/2$ and no bias voltage applied, such that for $\delta T > 0$ we have $T_\tf{hot} = T_L$ and $T_\tf{cold} = T_R$. When 
$\tc{W}_\tf{ext}>0$ we have a motor device driven by a heat current in response to a thermal gradient (heat engine), then the input power should be 
given by $-Q_{J_\tf{hot}}/ \tau$, and Equation~(\ref{eq:heat-motor}) implies
\begin{equation}
\eta = -\frac{\tc{W}_\tf{ext}}{Q_{J_\tf{hot}}} = 1 + \frac{Q_{J_\tf{cold}}}{Q_{J_\tf{hot}}} \leq 1-\frac{T_\tf{cold}}{T_\tf{hot}} = \eta_\tf{carnot}. \label{eq:hh-m}
\end{equation}
As compared with Equation~(\ref{eq:hc-m}), we can see that the efficiency of the quantum heat engine, now given by
\begin{equation}
\eta = -\frac{\tc{C}_\tf{ext}}{\tc{C}_{J_\tf{hot}}^{(0)} \dot{\theta}^{-1} + \tc{C}_{J_\tf{hot}}^{(1)} + \tc{C}_{J_\tf{hot}}^{(2)} \dot{\theta}},
\end{equation}
is defined in the same range for $\tc{C}_\tf{ext}$ as in the electric motor. Regarding Figure~\ref{fig:4}b, the engine's normalized efficiency 
$\eta / \eta_\tf{carnot}$ looks similar to that of the electric motor. Perhaps the only difference here is that for the smallest orbit 
$\delta_\epsilon = 10 k_\tf{B} T$ the efficiency maximum is very low, such that it can not be appreciated on the employed scale of the plot. 

\begin{figure}[ht]
\begin{center}
 \includegraphics[width=0.5\textwidth]{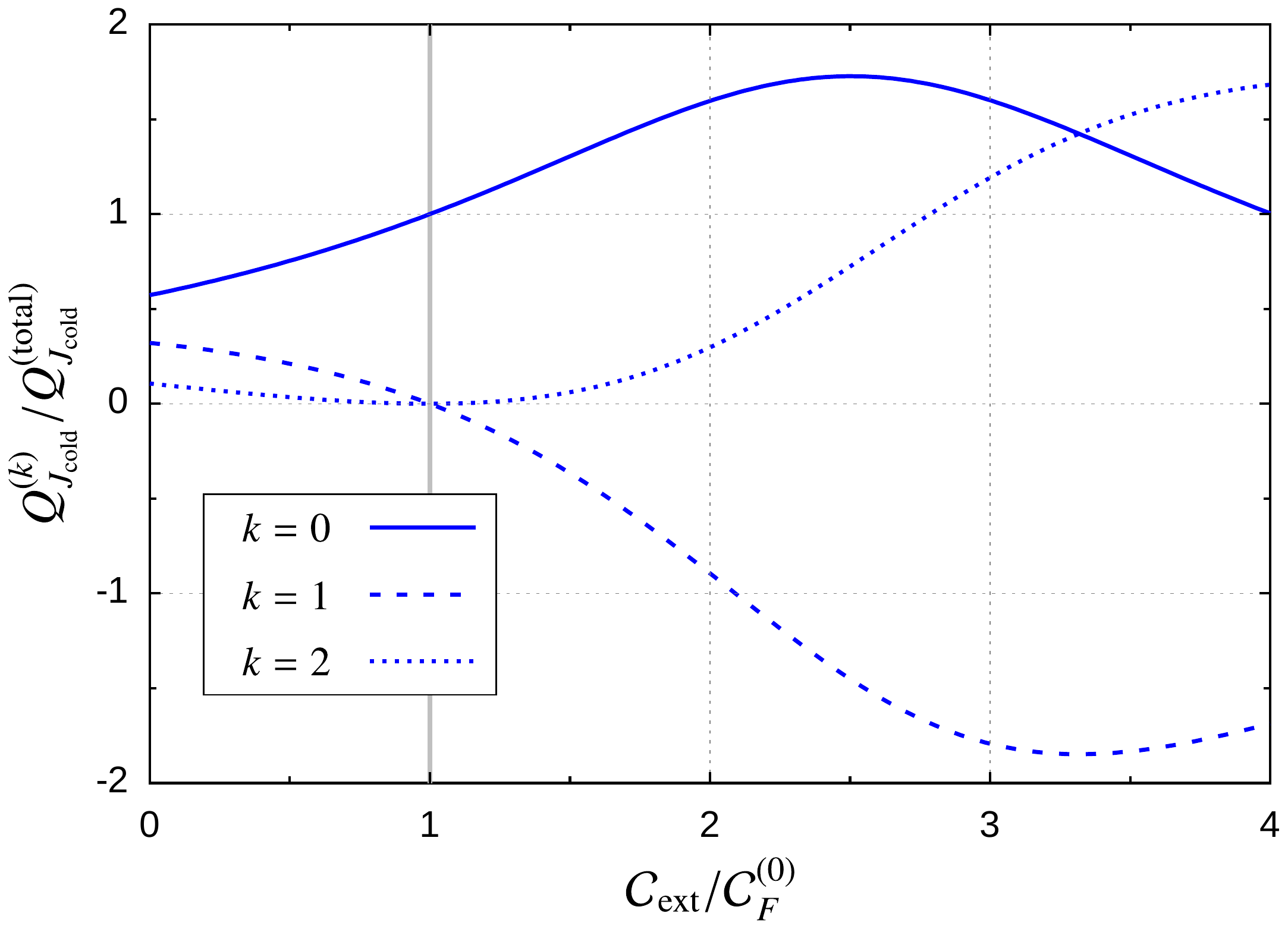}
\end{center}
\caption{
\new{
Different order contributions to the pumped heat as a function of the normalized external work for the case $\delta_\epsilon = 20 \, k_\tf{B}T$ (solid blue line in Figure~\ref{fig:4}b). Here $Q_J^{(\tf{total})}$ refers to the sum of all contributions.
}
}
\label{fig:6}
\end{figure}

Now we move to the heat pump region where the device acts as a refrigerator, as we demand that the heat current flows against the direction dictated by $\delta T$. Therefore, in addition to the $\tc{W}_\tf{ext} < 0$ condition, the overall amount of pumped heat in the cold reservoir should be 
negative, i.e., $Q_{J_\tf{cold}} < 0$. The efficiency of the refrigerator, or coefficient of performance (COP), is then given by
\new{
\begin{equation}
\tf{COP} = \frac{Q_{J_\tf{cold}}}{\tc{W}_\tf{ext}} = \frac{Q_{J_\tf{hot}}}{Q_J}-1 \leq \frac{T_\tf{cold}}{T_\tf{hot}-T_\tf{cold}} = \tf{COP}_\tf{carnot},
\end{equation}
}
where we can consistently expand $Q_{J_\tf{cold}}$ in terms of $\dot{\theta}$. In Figure~\ref{fig:4}b we show the lowest order contribution from 
Equation~(\ref{eq:vel1}) as the next-order calculation does not change significantly the efficiencies in the considered regimes of the parameters. Again, we can observe in Figure~\ref{fig:4}b a gap region where the device is frustrated since the work delivered by the rotor is not enough as to reverse the direction of the heat current.
One of the differences with the electric counterpart is that, for the refrigerator, the \textit{normalized} COP develops a maximum that is always smaller than that of the quantum heat engine, while the obtained efficiency maxima (motor and pump) for a fixed orbit in Figure~\ref{fig:4}a are very similar. Additionally, for the chosen value $\delta T = 0.5 \, T$ and small orbit radius, the device can only work as a heat engine, and the refrigerator can not be activated even if the \new{external} force is large, as happens for $\delta_\epsilon = 20 \, k_\tf{B}T$ (solid blue line). The reason for this relies on the competition between the different orders in the pumped heat $Q_{J_\tf{cold}}$: The reduction of the leakage pumped current, \new{$Q_{J_\tf{cold}}^{(0)}$}, by increasing the magnitude of $\dot\theta$ is accompanied by an increase of the second-order contribution, $Q_{J_\tf{cold}}^{(2)}$, such that the first order term, $Q_{J_\tf{cold}}^{(1)}$, may not be able to compensate these two and the requirement $Q_{J_\tf{cold}} < 0$ can not be fulfilled, \new{see Figure~\ref{fig:6}}.

\section{Summary}

Throughout this manuscript, we revisited some fundamental aspects related with the physics of quantum motors and pumps. Previous results based on the steady-state properties and energy conservation law were extended to deal with arbitrary nonequilibrium conditions in a systematic way. By considering the dynamics of the mechanical degrees of freedom through a Langevin equation, we were able to treat the motor/pump protocols on the same footing. This allowed us to describe the related energy transfer processes through a single parameter: the \new{external} force done on the local system.

In the steady-state regime, we treated in general terms the validity of the constant velocity assumption, Section~\ref{sec:steady_state}. For arbitrary orders of the nonadiabatic expansion in the CIFs, this was linked to the separation between the electronic and mechanical dynamic scales through a large moment of inertia.
We then performed a general expansion (in terms of nonequilibrium sources) of the energy fluxes that take part in the quantum transport problem. This enabled us to derive an order-by-order scheme for the energy conservation law, Equation~(\ref{eq:1stLaw_vect}). This Equation may be of help in recognizing the physical processes that enter at each order in the expansion, thereby providing a useful tool for the analysis of nonlinear effects. To illustrate this, we discussed the leading orders of the global expansion, and showed how different types of expansions of the energy fluxes change the expressions for the efficiency of quantum motors and pumps.

In Section~\ref{sec:Coulomb}, we introduced a specific example of a quantum motor/pump based on a double quantum dot. There, we discussed in more depth how higher-order terms of the CIFs affect the stationary state conditions. We found that multiple solutions for the device's terminal velocity could in principle be available for a fixed choice of parameters (voltage and temperature biases, and \new{external} force). In such a case, the stability of such solutions imposes an additional constraint on the force coefficients, and the final steady state strongly depends on initial conditions. Interestingly, it is possible to obtain more than one stable solution, each of them belonging to a different operation mode of the device.
The treated example is also appealing as it is possible to study the transition between different operational modes by continuously moving the \new{external} force.
This corresponds to the point in which the steady-state velocity changes its sign and, with it, the direction of the energy flow.
When considering a specific type of pumped currents (charge or heat), there is an intermediate region where the pumping mechanism is ``frustrated''.
In this situation, the energy delivered by the \new{external} force is not enough as to reverse the natural direction of the charge or heat currents.
We found other interesting features of the studied example such as negative friction coefficients at finite voltages or a definite parity of the expansion coefficients with respect to the bias voltage and the temperature gradient, which is a manifestation of the inversion symmetry in the total energy flux.
We also used this example to numerically confirm the order-by-order energy conservation law up to third order in the final velocity.
Finally, for heat currents, we found parameter conditions under which the device can never work as a ``refrigerator'', even for large values of the \new{external} force.
We explained this behavior in terms of the competition between the different orders that participate in the pumped heat of the cold reservoir, highlighting the importance of the order-by-order conservation laws.

\vspace{0.5cm}

\noindent
\textit{Acknowlegdments.--} This work was supported by Consejo Nacional de Investigaciones Cient\'ificas y 
T\'ecnicas (CONICET), Secretar\'ia de Ciencia y Tecnolog\'ia -- Universidad Nacional de C\'ordoba (SECYT--UNC). All authors are members of CONICET.

\bibliographystyle{apsrev4-1_title}
\bibliography{cite}

\end{document}